\documentclass[11pt]{article}
\pdfoutput=1
\usepackage[margin=1.0in]{geometry}
\usepackage{amsmath,amssymb,graphicx}
\usepackage{hyperref}
\usepackage{slashed}
\usepackage{wasysym}
\usepackage{hhline}

\usepackage[utf8]{inputenc}

\usepackage[greek,english]{babel}

\usepackage[caption=false]{subfig}
\usepackage{booktabs}
\usepackage{soul}

\usepackage[font=small,labelfont=bf]{caption}
\usepackage{cite}
\usepackage{enumerate}

\newcommand{\jQCDs}{{\rm ISR}}

\newcommand{\be}{\begin{equation}}
\newcommand{\ee}{\end{equation}}
\newcommand{\beq}{\begin{equation}}
\newcommand{\eeq}{\end{equation}}

\newcommand{\beqn}{\begin{eqnarray}}
\newcommand{\eeqn}{\end{eqnarray}}
\newcommand{\bea}{\begin{eqnarray}}
\newcommand{\eea}{\end{eqnarray}}

\newcommand{\wino}{\widetilde W}

\usepackage[table]{xcolor}
\usepackage{bm}

\definecolor{gray}{cmyk}{0,0,0,0.05}
\newcolumntype{a}{>{\columncolor{gray}} l}

\newcommand{\met}{E_T^{\rm miss}}

\numberwithin{equation}{section}

\newcommand{\bino}{\widetilde{B}}

\usepackage{soul}

\begin{document}

%\begin{titlepage}

\begin{center}
%\hfill \\  %% Preprint Number
%\hfill \\

%PREPRINT NUMBER IFIC: IFIC/21-05

~
\vskip 1cm

{\LARGE \bf 
Monojet signatures from gluino and squark decays }

\vskip 1.5cm

{\large 
I\~naki Lara$^{(a)}$,
Trygve Buanes$^{(b)}$,
Rafa\l{} Mase\l{}ek$^{(a)}$, 
\\
\vspace{3mm}
Mihoko M.~Nojiri$^{(c,d,e)}$,
Krzysztof Rolbiecki$^{(a)}$ and
Kazuki Sakurai$^{(a)}$
}

\vskip 1cm

$^{(a)}${\em
Institute of Theoretical Physics, Faculty of Physics,\\
University of Warsaw, ul.~Pasteura 5, PL-02-093 Warsaw, Poland %\\[0.1cm]
}

\vskip 0.3cm

$^{(b)}${\em
%Western Norway University of Applied Sciences, Bergen, Norway
Western Norway University of Applied Sciences, Inndalsveien 28, Bergen, Norway
}

\vskip 0.3cm

$^{(c)}${\em
Theory Center, IPNS, KEK, 1-1 Oho,
Tsukuba, Ibaraki 305-0801, Japan}

\vskip 0.3cm

$^{(d)}${\em
The Graduate University of Advanced Studies (Sokendai),
1-1 Oho, \\ Tsukuba, Ibaraki 305-0801, Japan}

\vskip 0.3cm

$^{(e)}${\em
Kavli IPMU (WPI), University of Tokyo,
5-1-5 Kashiwanoha, Kashiwa, Chiba 277-8583, Japan}

\end{center}

%\vskip 0.5cm
%~
\vskip 1.5cm

%\today
\begin{abstract}

We study the monojet and dijet channels at the LHC as a tool for searching for squarks and gluinos.
We consider two separate 
R-parity conserving supersymmetric
scenarios.
In the first scenario 
we postulate a large mass hierarchy between squarks 
($\tilde q$) and winos ($\widetilde W$), 
and wino-like neutralino 
is assumed to be the lightest supersymmetric particle (LSP).
The associated squark-wino production, $pp \to \tilde q \widetilde W$, 
then leads to a monojet-like signature,
where the high $p_T$ jet is originated from the squark decay, 
$\tilde q \to q + \widetilde W$. 
We demonstrate 
that this associated production, 
as well as the $pp \to \widetilde W \widetilde W + {\rm jets}$ production,
have a significant impact on the
exclusion limit in the squark-neutralino mass plane.
The second scenario postulates that the lighter of the squark and gluino is only a few GeV heavier than the LSP neutralino.
The associated squark-gluino production, $pp \to \tilde q \tilde g$, then leads to 
a distinctive monojet signature,
where the high $p_T$ jet is produced from the decay of the heavier coloured 
particle into the lighter one ($\tilde q \to q + \tilde g$ for $m_{\tilde q} > m_{\tilde g}$
and $\tilde g \to q + \tilde q$ for $m_{\tilde g} > m_{\tilde q}$).
The lighter coloured particle is 
effectively regarded as an invisible particle 
since the decay products are soft
due to the approximate mass degeneracy.
We recast existing monojet and dijet analyses and find a non-trivial exclusion limit 
in the squark-gluino mass plane in this scenario.

\end{abstract}

%\end{titlepage}

\vskip 1.5cm

\newpage
%%%%%%%%%%%%%%%%%%%%%%%%%%%%%%%%%%%%%%%%%
\section{Introduction}
\label{sec:intro}
%%%%%%%%%%%%%%%%%%%%%%%%%%%%%%%%%%%%%%%%%

Observed evidence of dark matter in the Universe strongly indicates the existence of a stable and neutral particle, $\chi$, beyond the Standard Model (SM) particle content.
In a scenario where the stability of the dark matter is guaranteed by a $Z_2$ symmetry, e.g.\ $R$-parity in supersymmetry (SUSY) \cite{Farrar:1978xj,Weinberg:1982tp,Ellis:1983ew}, high energy hadron colliders may produce the dark matter particles in pairs, associated with a few high $p_T$ jets originated from initial state QCD radiation (ISR); $pp \to \chi \chi + \jQCDs$.
A particularly useful channel to detect such events is called {\it monojet} \cite{Alwall:2008va},
which is defined as a type of event selection that requires no isolated lepton, a small number of high $p_T$ jets (usually up to four)\footnote{
%Despite the name, ``monojet'' channels typically allow more than one jet. However first searches indeed required exactly one energetic jet~\cite{ATLAS-CONF-2011-096} or at most two~\cite{CMS-PAS-EXO-11-059}, this requirement was later relaxed to improve sensitivities~\cite{Dreiner:2012gx}.
Despite the name, ``monojet'' channels typically allow more than one jet as in \cite{ATLAS:2021kxv,CMS:2021far}, %\st{Historically the first analysis }\cite{ATLAS-CONF-2011-096} \st{required one energetic jet and vetoed the second leading jet with some $p_T$ threshold. 
%or at most two~\cite{CMS-PAS-EXO-11-059}, The requirement of second jet veto was later removed to improve sensitivities}~\cite{CMS-PAS-EXO-11-059, Dreiner:2012gx}  
though historically a lower jet multiplicity was imposed \cite{ATLAS-CONF-2011-096, CMS-PAS-EXO-11-059, Dreiner:2012gx} with variable veto threshold on additional jest. } and
a large missing transverse momentum, ${\bf p}_T^{\rm miss}$, recoiling against the high $p_T$ jets.\footnote{We adopt a notation used by ATLAS;  
${\bf p}_T^{\rm miss} = (p_T^{{\rm miss},x}, p_T^{{\rm miss},y})$ denotes the two-component missing transverse momentum 
and $\met \equiv |{\bf p}_T^{\rm miss}|$.} 
After imposing a tight cut on the $\met$ and $p_T$ of the jets, 
the background is dominated by
$pp \to Z + \jQCDs$, followed by $Z \to \nu \bar \nu$, which
amounts to $60 - 80\%$ of the total SM background depending on the cuts.
The second largest background,
$pp \to W^\pm + \jQCDs$, followed by $W^\pm \to \tau^\pm \nu (\bar \nu)$,
takes up $\sim 10\%$ of the total
\cite{ATLAS:2021kxv,CMS:2021far}. In a similar spirit, the ATLAS multijet search~\cite{ATLAS:2020syg} contains a subset of signal regions where the number of final state jets is constrained to 2--3, however with an additional requirement that both leading jets are very hard, i.e.\ $p_T > 250$~GeV. 

Recently ATLAS \cite{ATLAS:2021kxv} and CMS \cite{CMS:2021far} have analysed Run-2 data in monojet channels. In addition to the conventional direct dark matter production scenarios mentioned above, ATLAS interpreted data for the squark pair production associated with hard QCD radiation, $pp \to \tilde q \tilde q + \jQCDs$, postulating the mass of the lightest, bino-like neutralino $\tilde \chi_1^0$, which is assumed to be
the lightest supersymmetric particle (LSP) and stable, is only a few tens of GeV smaller than the squark mass.
In this case, 
squark may be treated as an invisible particle,
since its decay, $\tilde q \to q \tilde \chi_1^0$,
is soft and the squark passes its momentum almost entirely to the neutralino.
The only high $p_T$ visible objects in the event 
are $\jQCDs$,
and a large $\met$ is generated to balance 
the $\jQCDs$.
Such events dominantly contribute to the monojet channel.
Observing no excess in the signal region, ATLAS has placed a limit on the squark mass $m_{\tilde q} \gtrsim 800 $ GeV,
depending on the mass difference $m_{\tilde q} - m_{\tilde \chi_1^0} \sim \mathcal{O}(10)$ GeV.
This example demonstrates that the monojet channel may be a powerful tool
to look for coloured particles 
in the compressed mass region.

A common feature of the above two signal processes
\begin{description}
%\begin{center}
\item[~~~~~~~~~~(i)]
$pp \to \chi \chi + \jQCDs$
\item[~~~~~~~~~\,(ii)]
$pp \to \tilde q \tilde q + \jQCDs$
%\end{center}
\end{description}
is that the high $p_T$ jets (and large $\met$) have their origin in QCD radiation.
This is challenging from  the analysis point of view, since 
QCD radiation has a monotonically falling spectrum and the signal acceptance becomes low once tight $p_T$ cuts are imposed on the jets.
Furthermore, 
in the background process,
$pp \to Z(W^{\pm}) + \jQCDs$,
the jets are also originated from QCD radiation.
Thus, their kinematical distributions are similar, which leads to a poor signal-background separation.
Nevertheless, monojet channels are commonly used as a powerful tool to constrain this type of signal processes.

In this study, we point out that  the mono- and di-jet channels are also sensitive to the following processes:
% signature  of $pp\to\tilde q \tilde \chi$ with large squark-electroweakino mass splitting 
% and  $pp\to\tilde q \tilde g$ with the mass degenerate neutralino LSP, $m_{\tilde \chi} \sim {\rm min}(m_{\tilde q}, m_{\tilde g})$.
% We show that by recasting mono- and di-jet searches, the limits on sparticle masses are significantly improved with respect to the current ATLAS and CMS results.
% Precisely speaking, we consider the following two SUSY scenarios:
%
\begin{description}
\item[~~~~~~(I)]~ $pp \to \tilde q \tilde \chi$,~ followed by~ $\tilde q \to q + \tilde \chi$ ~~$\cdots$~ (\, a few TeV $\gtrsim m_{\tilde q} \gg m_{\tilde \chi}$ \,)
\item[~~~~\,\,(II)]~ $pp \to \tilde q \tilde g$,~ followed by~
$
\left\{
\begin{array}{ll}
\tilde q \to q + \tilde g & 
~~\cdots~(\, {\rm a \,\,few \,\, TeV} \gtrsim m_{\tilde q} \gg m_{\tilde g} = (1 + \epsilon)m_{\tilde \chi}\,)
\\ 
\tilde g \to q + \tilde q & 
~~\cdots~
(\, {\rm a \,\,few \,\, TeV} \gtrsim  
m_{\tilde g} \gg m_{\tilde q} = (1 + \epsilon)m_{\tilde \chi}\,)
\end{array}
\right.
$
\end{description}
with $0 < \epsilon \ll 1$.
Other than hard QCD radiation, 
the final states of {\bf (I)} and {\bf (II)} have a single high $p_T$ quark-jet (denoted by $q$) from the decay of a coloured SUSY particle.\footnote{Since neutralinos are colour-singlet, $SU(3)_C$ gauge interaction does not allow $\tilde g \to g \tilde \chi_1^0$ at tree level. This decay mode is generated at 1-loop level but the branching ratio is negligible compared to $\tilde g \to q \tilde q$. Therfore, we do not consider this decay mode.} Unlike $\jQCDs$ in the aforementioned processes {\bf (i)} and {\bf (ii)}, the $q$-jet has the energy scale characterised by the mass differences between the squark and electroweakino in ({\bf I}), and the squark and gluino in ({\bf II}). Kinematical distributions of $q$-jet are therefore different from those of $\jQCDs$ in the SM background, which may help to discriminate the signal from background in the analysis.
We recast existing mono- and di-jet searches and show in section \ref{sec:sq-wino} that the process ({\bf I}) has a large impact on sparticle mass limits in the squark-electroweakino simplified model, when included together with the ordinary pair production processes, $pp \to \tilde q \tilde q, \tilde \chi \tilde \chi$.
Similarly, in section \ref{sec:gl-sq}
we study the effect of the process ({\bf II}) in the gluino-squark-bino simplified model.
Including the process 
({\bf II}) together with the pair production processes, $pp \to \tilde q \tilde q, \tilde g \tilde g$, we derive the sparticle mass limit in 
the ($m_{\tilde g}$, $m_{\tilde q}$) plane.
% the limits on sparticle masses are significantly improved with respect to the current ATLAS and CMS results when the above processes are included together with the pair productions, $pp \to \tilde q \tilde q, \tilde g \tilde g, \tilde \chi \tilde \chi$. 
%We study the effect of processes 
%({\bf I}) and ({\bf II}) exclusively in sections \ref{sec:sq-wino}
%and \ref{sec:gl-sq}, respectively.

Regarding the process {\bf (II)}, we emphasise that
the associated production $pp \to \tilde q \tilde \chi$ has a larger cross section than for the squark pair production, $pp \to \tilde q \tilde q$, in the heavy-squark light-neutralino region (e.g.~$m_{\tilde q} \gtrsim 800$ GeV and $m_{\wino} \lesssim 200$ GeV for wino-like LSPs, $m_{\tilde q} \gtrsim 2$ TeV and $m_{\bino} \lesssim 100$ GeV for bino-like LSPs). 
This is because producing two heavy-squarks becomes energetically too expensive so that the single squark production $pp \to \tilde q \tilde \chi$ takes over, despite being partially induced by electroweak gauge interactions.
%  the main production, although 
% the latter is partially induced by electroweak gauge interactions.}
We also note that the current squark mass limit is already pushed to $\sim 1.8$ TeV
in a hierarchical mass region
and the limit is derived by considering only the squark pair production.
One therefore expects 
the associated production
may have a large impact on the squark mass limit.
We investigate this issue in this paper.

The rest of the paper is organised as follows.
In the next section, we study 
the associated $\tilde q$-$\tilde \chi$ production process {\bf (I)}.
%assuming $\tilde \chi$ is wino-like.
We compare the cross sections
of the associated production
$pp \to \tilde q \tilde \chi$
and the pair production 
$pp \to \tilde q \tilde q$
and demonstrate that the former 
may have larger cross section when
$m_{\tilde \chi} \ll m_{\tilde q}$.
After comparing the kinematical distributions
of these processes,
we recast the relevant ATLAS 
analyses employing mono- and di-jet event selections
\cite{ATLAS:2020syg, ATLAS:2021kxv}
assuming wino-like LSPs.
We show the exclusion limit in the squark-wino mass plane extends significantly when the squark-wino associated production is included in the signal sample.

In section \ref{sec:gl-sq} we study the monojet-like signature in the scenario where both squrks and gluino are light enough to be produced at the LHC but the lighter one is mass-degenerate with the bino-like LSP neutralino. Such a mass spectrum is motivated by the gluino-bino and squark-bino coannihilation scenarios \cite{DeSimone:2014qkh,Profumo:2004wk,Harigaya:2014dwa,Ellis:2015vaa}. We compare kinematical distributions of process {\bf (II)}, $pp \to \tilde q \tilde g$, with those of the pair productions, $pp \to \tilde q \tilde q ~(+ \jQCDs)$ and $pp \to \tilde g \tilde g ~(+ \jQCDs)$.
We estimate the current exclusion limit in the squark-gluino mass plane by recasting the relevant ATLAS analyses  \cite{ATLAS:2021kxv, ATLAS:2020syg} assuming that the bino-like neutralino is almost mass degenerate with the lighter coloured SUSY particle. Section \ref{sec:concl} is devoted to the conclusions.

%%%%%%%%%%%%%%%%%%%%%%%%%%%%%%%%%%%%%%%%%
\section{Monojet from squark-wino productions}
\label{sec:sq-wino}
%%%%%%%%%%%%%%%%%%%%%%%%%%%%%%%%%%%%%%%%%

If low energy supersymmetry is realised in nature,
high energy hadron-hadron colliders should be able to produce squarks, depending on the squark mass and the collider energy.
Usually, results of squark searches are interpreted in the squark-neutralino simplified model with decoupled gluino and  
the exclusion limits are presented in the ($m_{\tilde q}, m_{\tilde \chi_1^0}$) plane \cite{ATLAS:2020syg, CMS:2019zmd}.  
So far, ATLAS and CMS have included only the squark pair production,
\be \label{eq:QCD}
pp \to \tilde q  \tilde q^*\,,
\ee
in their analyses and 
the
associated squark-electroweakino production, 
\be
pp \to \tilde q  \tilde \chi\,,
\label{eq:QW}
\ee
has been omitted.
Indeed, the former is a pure QCD process ($\sigma \propto \alpha_s^2$) and 
has much larger cross section than the latter
($\sigma \propto \alpha_s \alpha_W$)
for $m_{\tilde q} \sim m_{\tilde \chi}$. We note, however, that the electroweak process of Eq.~\eqref{eq:QW} is only negligible when the $\tilde \chi$s are higgsino-like. In case of winos (and binos to some extent), the contribution cannot be neglected, as we shall argue in this sectione following.  

The cross section of the  production \eqref{eq:QCD} decreases quickly as the squark mass increases, since two squarks are produced.
This means for a hierarchical mass spectrum, $m_{\tilde q} \gg m_{\tilde \chi}$,
the latter process may become more important than the former.
The current squark mass limit with Run-2 data has already been pushed to $\sim 1.8$ TeV 
for $m_{\tilde \chi} \lesssim 300$ GeV \cite{ATLAS:2020syg, CMS:2019zmd}.
It is therefore urgent to study the effect of the squark-electrowino associated production \eqref{eq:QW}
in the squark searches.

%%%%%%%%%%%%%%%%%%%%%%%%%%%%%%%%%%%%%%%%%
\subsection{The production cross sections}
\label{sec:cross section}
%%%%%%%%%%%%%%%%%%%%%%%%%%%%%%%%%%%%%%%%%

\begin{figure}[t!]
\centering
      \includegraphics[width=0.33\textwidth]{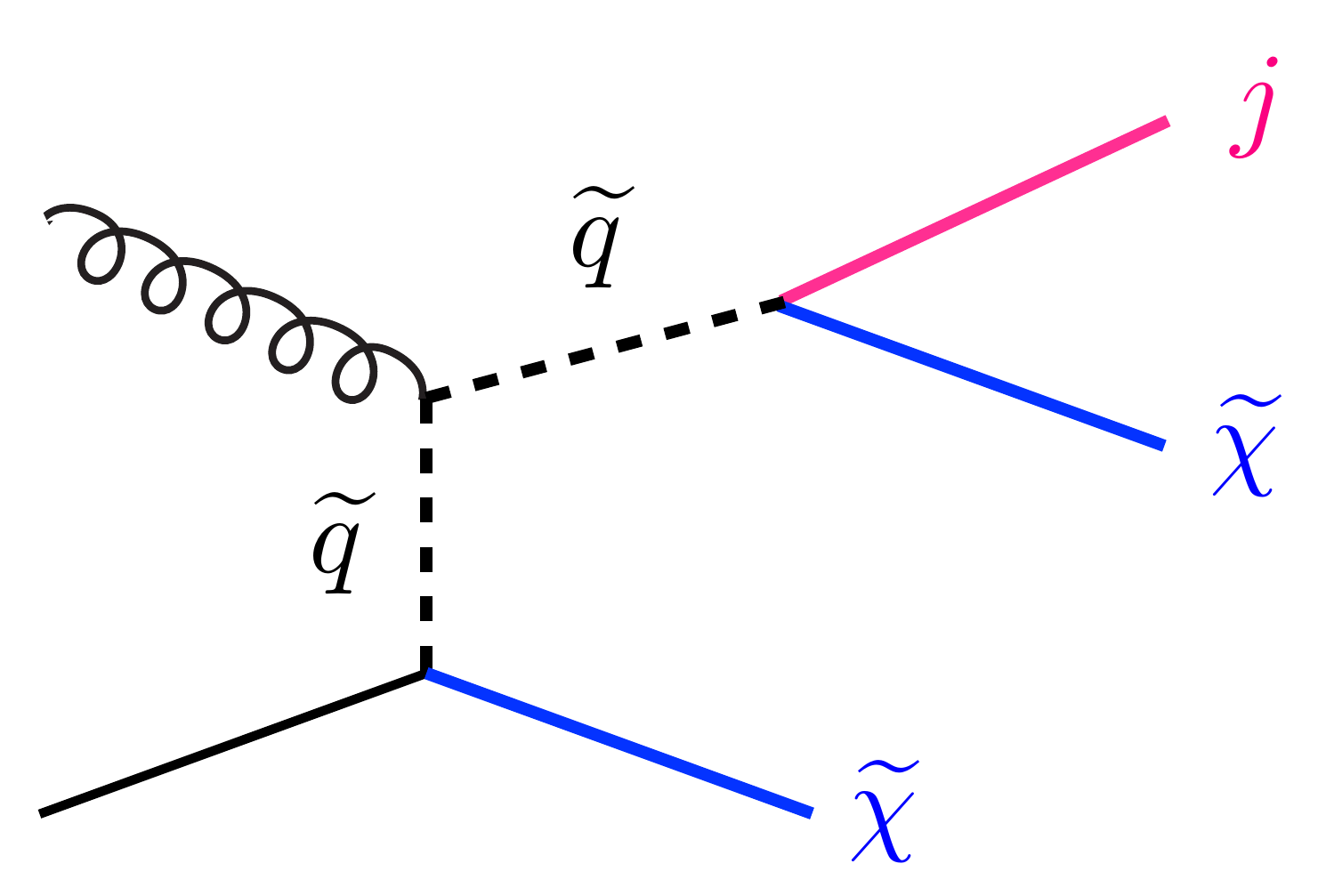}
     
\caption{A diagram for squark-electroweakino productions.
}
\label{fig:QN_diagram}
\end{figure}

We assume a large mass gap between the squark, $\tilde q$, and the electroweakino, $\tilde \chi$, keeping the squark mass still within the LHC reach, i.e.\ (a few TeV) $> m_{\tilde q} \gg m_{\tilde \chi}$. Gluinos are taken to be decoupled so that the produced squarks subsequently decay into a quark and an electroweakino, $\tilde q \to q + \tilde \chi$. In this case, the final state is given by the two electroweakinos and a single high-$p_T$ jet originated from the squark decay, depicted in Fig.~\ref{fig:QN_diagram}. If $\tilde \chi$ is stable and invisible in the detector, the process contributes to the monojet and dijet channels. The situation is trivially realised by identifying $\tilde \chi$ to be a bino-like LSP, but it can be effectively realised also for the wino- and higgsino-like LSP scenarios. In the latter cases, $\tilde \chi$ should be identified as a triplet ($\tilde \chi_1^\pm, \tilde \chi_1^0$)~= ($\widetilde W^\pm, \widetilde W^0$) or a pair of doublets $(\tilde \chi_{1}^\pm, \tilde \chi_{1,2}^0)$~=$(\widetilde H^{\pm}_{u/d}, \widetilde H^{0}_{u/d})$ of the $SU(2)_L$
for the wino- and higgsino-like LSP scenarios, respectively. Thus from the decay point of view the three assumptions are equivalent and experimentally indistinguishable.
The particles within the same multiplet are almost mass degenerate and 
the decays of the heavier to the lighter one within the multiplet do not leave decay products above the kinematical threshold.  
An example is a decay of the charged wino into the neutral wino, $\tilde \chi_1^+ 
\to \tilde \chi_1^0 + X_{\rm soft}$,
the decay product,
$X_{\rm soft}$, are very soft and not to be reconstructed as a signal candidate.
Since heavier particles in the multiplet eventually decay into the neutral LSP, $\tilde \chi_1^0$, 
the multiplet as a whole can effectively be treated as a missing particle.\footnote{If the charged wino is heavily boosted, its decay products, either leptons or mesons, might eventually get enough transverse momentum and become identified in the detector. 
In such a case, the event may be rejected
by a lepton veto or a cut demanding large separation between the missing transverse momentum 
and subleading jets. 
We have checked, however, that for small mass differences, $\Delta m = m_{\tilde W^\pm} - m_{\tilde W^0} < 1$ GeV, the exclusion limit is insensitive to $\Delta m$.} 

As is evident from the diagram in Fig.~\ref{fig:QN_diagram}, the amplitude of the $g q \to \tilde q \tilde \chi$ process is proportional to the $SU(3)_C$ gauge coupling and the Yukawa coupling of $\tilde q$-$\tilde \chi$-$q$ interaction. If $\tilde \chi$ is higgsino-like, this Yukawa coupling is proportional to the mass of the quark in the initial state, which makes the cross section too small for the process to be observed. For bino- and wino-like $\tilde \chi$, the Yukawa coupling is given by the corresponding gauge couplings, $g_Y$ and $g_W$, respectively. The cross section is the largest when $\tilde \chi$ is wino-like since $g_W > g_Y$ and the multiple final states, $\tilde q \widetilde W^\pm$ and $\tilde q \widetilde W^0$, contribute. In the wino-like LSP scenario, the charged wino may be long-lived, $c \tau_{\tilde W^\pm} \sim {\cal O}(1)$\,cm, if $m_{\tilde \chi_1^\pm} - m_{\tilde \chi_1^0} \lesssim 150$ MeV. In this case, the model is constrained by the searches looking for disappearing tracks~\cite{ATLAS:2022rme}. In this paper, we do not consider such an exotic signature and assume that the chargino lifetime is short enough to evade the disappearing track constraint, that is $m_{\tilde \chi_1^\pm} - m_{\tilde \chi_1^0} \gtrsim 400$ MeV. The winos up to 160 GeV are however constrained~\cite{Buanes:2022wgm} by the jet+$\met$ search ~\cite{ATLAS:2020syg}. The effect is taken into account also in this study.

\begin{figure}[t!]
    \centering
    \includegraphics[width=0.6\textwidth]{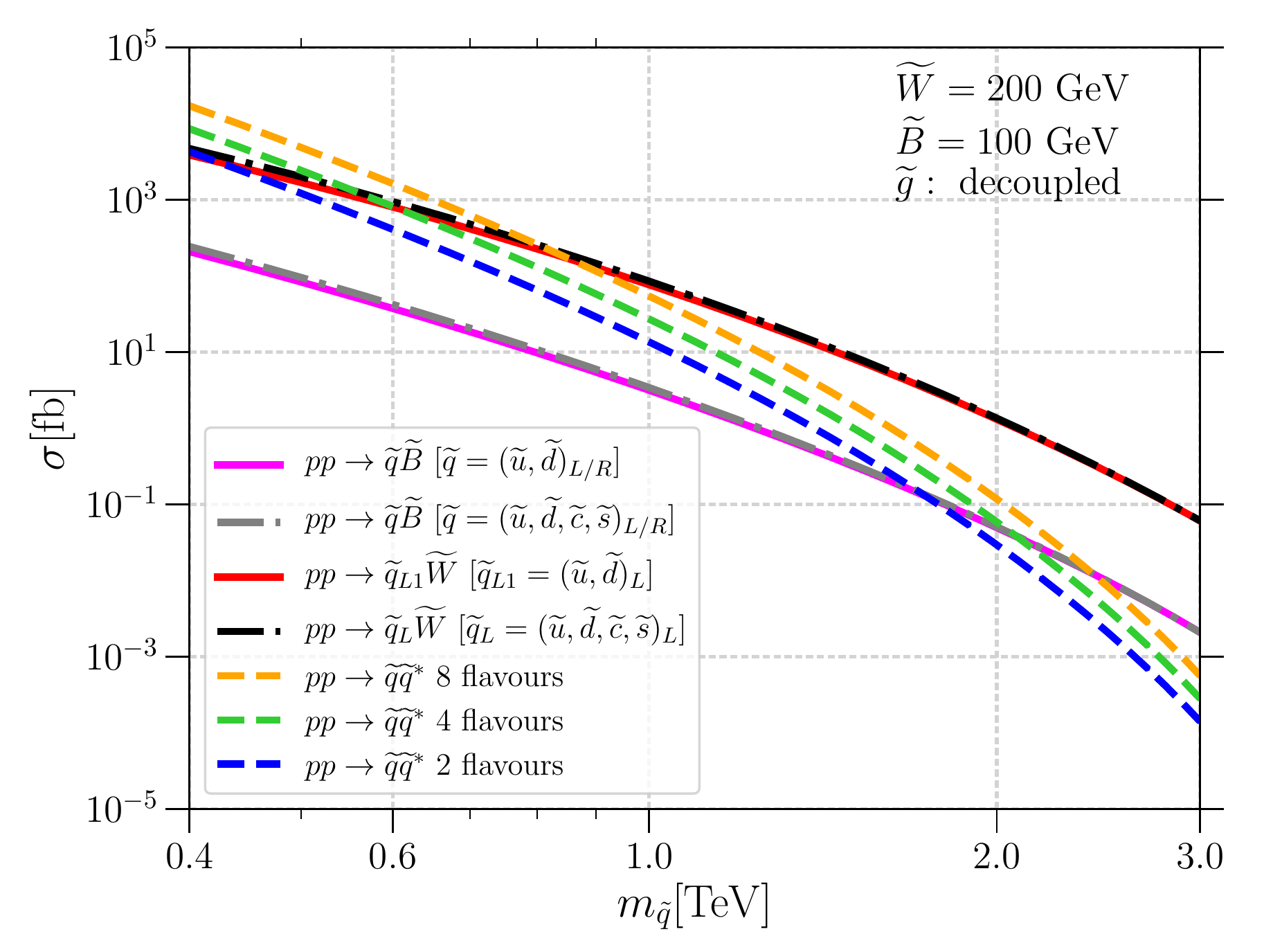}
    \caption{The cross sections for various production channels as a function of the produced squark mass.}
    \label{fig:QN_xsec}
\end{figure}

Fig.~\ref{fig:QN_xsec} displays the cross sections for various production channels as a function of the produced squark mass. 
We include squark-bino associated production here for experimental comparison since models with bino are typically used by collaborations to present exclusion limits.
For the associated squark-bino and squark-wino production modes,
{\tt Resummino 3.1.1}
\cite{Fuks:2013vua,Fiaschi:2022odp}
is used to calculate
the NLO cross sections,
fixing the wino (bino) mass at 200 (100) GeV.
For the squark pair productions,
the plot shows the NNLO+NNLL cross sections calculated by the {\tt SUSY cross section Working Group} \cite{Beenakker:1996ch,Kulesza:2008jb,Kulesza:2009kq,Beenakker:2009ha,Beenakker:2011sf,Beenakker:2013mva,Beenakker:2014sma,Beenakker:2016lwe} assuming decoupled gluino.

The three dashed lines in Fig.~\ref{fig:QN_xsec} represent the cross sections, $\sum_{i=1}^{N_f} \sigma(pp \to \tilde q_i \tilde q^*_i)$, where the final state includes $N_f = 8$ (orange), 4 (green) and 2 (blue) squark flavours.
The  cross section
does not depend on the squark-types included in the sum 
since the production mechanism is dominated by the $s$-channel gluon exchange and the gluon-gluon fusion
in the $\tilde g$ decoupling limit.

The squark-bino cross sections,
$\sum_i \sigma (pp \to \tilde q_i \widetilde B)$, for the 1-generation, $(\tilde u, \tilde d)_{L/R}$ [$N_f = 4$],
and 2-generation, 
$(\tilde u, \tilde d, \tilde c, \tilde s)_{L/R}$ [$N_f = 8$],
scenarios are represented by the solid-magenta and dotted-dashed-grey curves, respectively.
The two curves are almost on top of each other because 
adding the second generation only mildly increases the production rate since the parton distribution functions (PDFs) for the second generation quarks ($c$, $s$) are much smaller than those for the first generation ($u$, $d$).
Comparing them with the squark pair productions, we observe that for $m_{\tilde q} \gtrsim 2$ TeV, the squark-bino cross section surpasses the squark-squark one for the 1-generation [$N_f=4$] scenario, and  $m_{\tilde q} \gtrsim 2.4$ TeV for the 2-generation [$N_f = 8$] scenario.

Unlike the squark-bino production,
only the left-type squarks, $\tilde q_L$, contribute
to the squark-wino associated production. 
We therefore show in Fig.~\ref{fig:QN_xsec}
$\sigma (pp \to \tilde q_{L1} \widetilde W)$ with $q_{L1} = (\tilde u, \tilde d)_L$
and
$\sigma (pp \to \tilde q_{L} \widetilde W)$ with $q_{L} = (\tilde u, \tilde d, \tilde c, \tilde s)_L$
by the solid-red and dotted-dashed-black lines, respectively.
Similarly to the squark-bino case,
the two lines are almost overlapped since
the squark-wino cross section for the second generation 
is small compared to that for the first generation. 
The solid-red line, i.e.~$\sigma (pp \to \tilde q_{L1} \widetilde W)$,
may be compared with 
the squark pair production with $N_f=2$ if all squarks other than $(\tilde u, \tilde d)_L$ are decoupled,
or $N_f=4$ if both left- and right-type first generation squarks, $(\tilde u, \tilde d)_{L/R}$, remain light.
In the former (latter) case,
the squark-wino cross section becomes larger than squark-squark cross section for $m_{\tilde q} \gtrsim 500$ (600) GeV.
Similarly,
the dotted-black line, i.e.~$\sigma (pp \to \tilde q_{L} \widetilde W)$, should be confronted with the $N_f=4$ (8) squark-squark cross section for the case where only left-type squarks  
(all first and second generation squarks) are light.
We see that $\sigma (pp \to \tilde q_{L} \widetilde W)$
surpasses the squark-squark cross section at $m_{\tilde q} \simeq 600$
(800) GeV for $N_f = 4$ (8).
Since the current squark mass limit is around 1.3, 1.6 and 1.8 TeV for $N_f = 2$, 4 and 8, respectively, for the bino-like LSP scenario with $m_{\tilde q} \gg m_{\tilde \chi_1^0}$, this implies that the inclusion of the squark-wino associated production is important when calculating the squark mass limit on the wino-like LSP scenario.    

In the rest of this section, we focus on the wino-like LSP scenario and consider the following cases:
\begin{center}
\begin{tabular}{ c l }
          & \textbf{Light squarks} 
\\ 
\textbf{2-flavour} & $\tilde u_L, \tilde d_L$
\\  
\textbf{4-flavour} & 
$\tilde u_L, \tilde d_L, \tilde c_L, \tilde s_L$
\\  
\textbf{8-flavour} & 
$\tilde u_L, \tilde d_L, \tilde c_L, \tilde s_L,
\tilde u_R, \tilde d_R, \tilde c_R, \tilde s_R$
\end{tabular}
\end{center}
In each case, 
squarks other than listed as ``light'' are very heavy and decouple.

\begin{figure}[t!]
    \centering
    \includegraphics[width=0.49\textwidth]{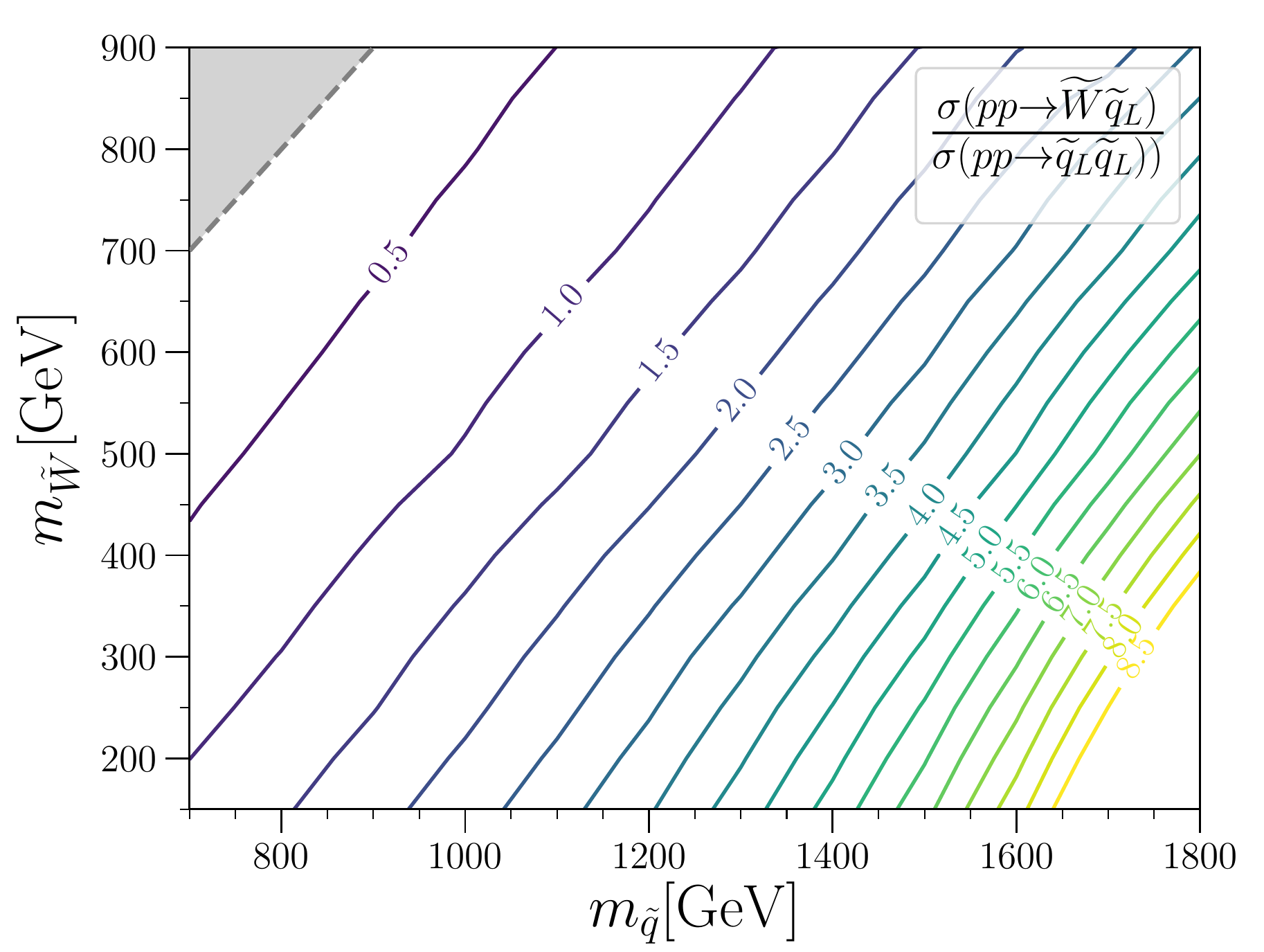}
    \includegraphics[width=0.49\textwidth]{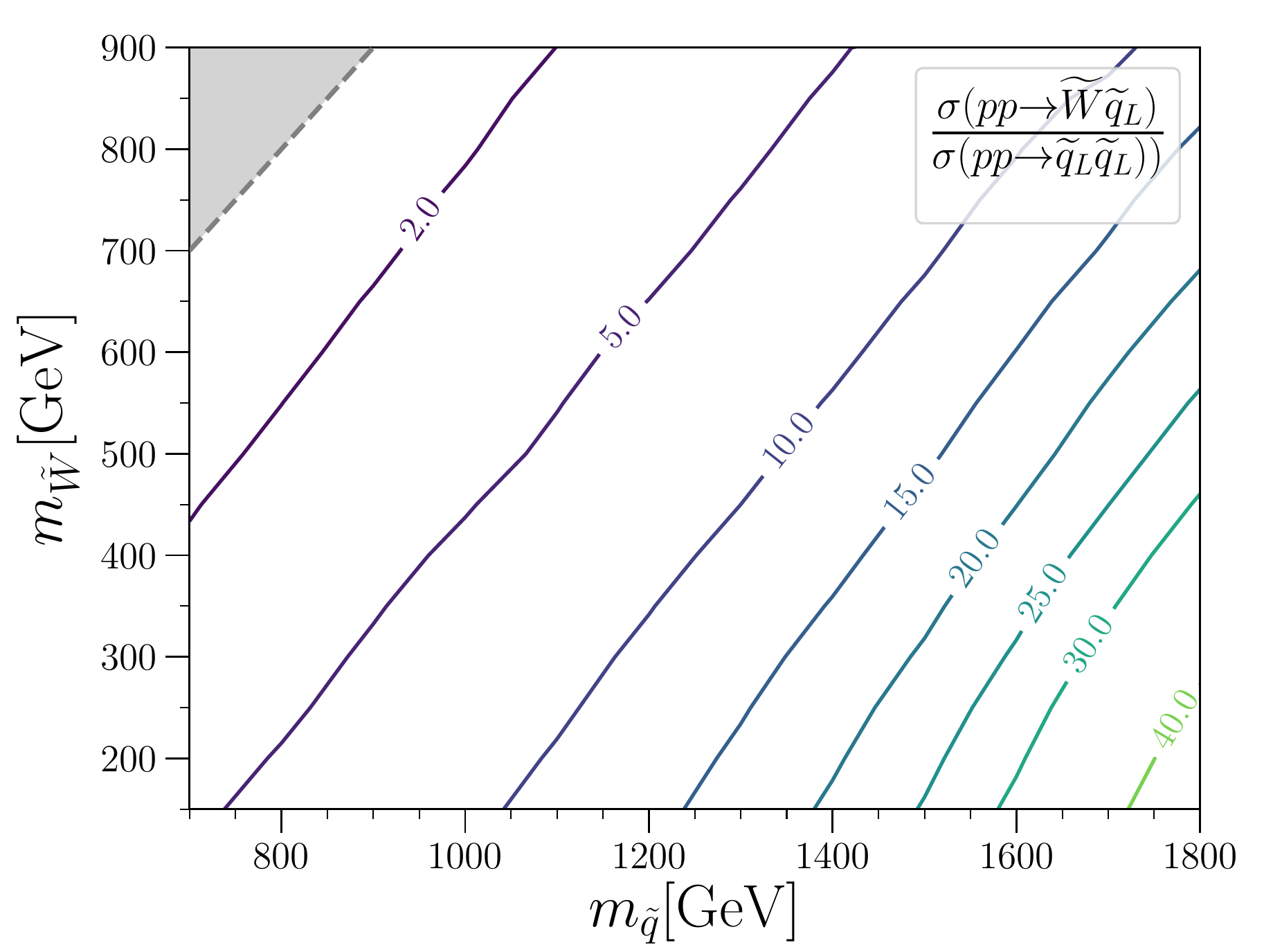}
    \caption{The cross section ratio between 
    $pp \to \tilde q_L \widetilde W$
    and
    $pp \to \tilde q \tilde q^*$ processes.
    The left and right plots correspond to
    the 8- and 2-flavour scenarios, respectively. 
    }
    \label{fig:xsec_QN_planes_ratio}
\end{figure}

In Fig.~\ref{fig:xsec_QN_planes_ratio} we show the cross section ratio $\sigma(pp \to \tilde q_L \widetilde W)/\sigma(pp \to \tilde q \tilde q^*)$ for the 8- and 2-flavour scenarios in the left and right panels, respectively. We see that the relative importance of the associated production is significant for larger $m_{\tilde q}$ and smaller $m_{\widetilde W}$. For the 8-flavour case,  the cross section of the associated production is already 2 times larger than that of the squark pair production  at $(m_{\tilde q}$, $m_{\widetilde W}) = (1000, 250)$\,GeV. For the 1.5 TeV squarks,  $\sigma(pp \to \tilde q_L \widetilde W)/\sigma(pp \to \tilde q \tilde q^*) \simeq 6$ (2) for $m_{\widetilde W} = 200$ (800) GeV. For the 2-flavour scenario, the effect is more enhanced. At $m_{\tilde q} = 1$ TeV, $\sigma(pp \to \tilde q_L \widetilde W)/\sigma(pp \to \tilde q \tilde q^*) \simeq 7$ (2) for $m_{\widetilde W} = 200$ (800) GeV, and  for $m_{\tilde q} = 1.5$ TeV the ratio is 25 (7) for   $m_{\widetilde W} = 200$ (800) GeV, respectively.

\begin{figure}[t!]
    \centering
    \includegraphics[width=0.49\textwidth]{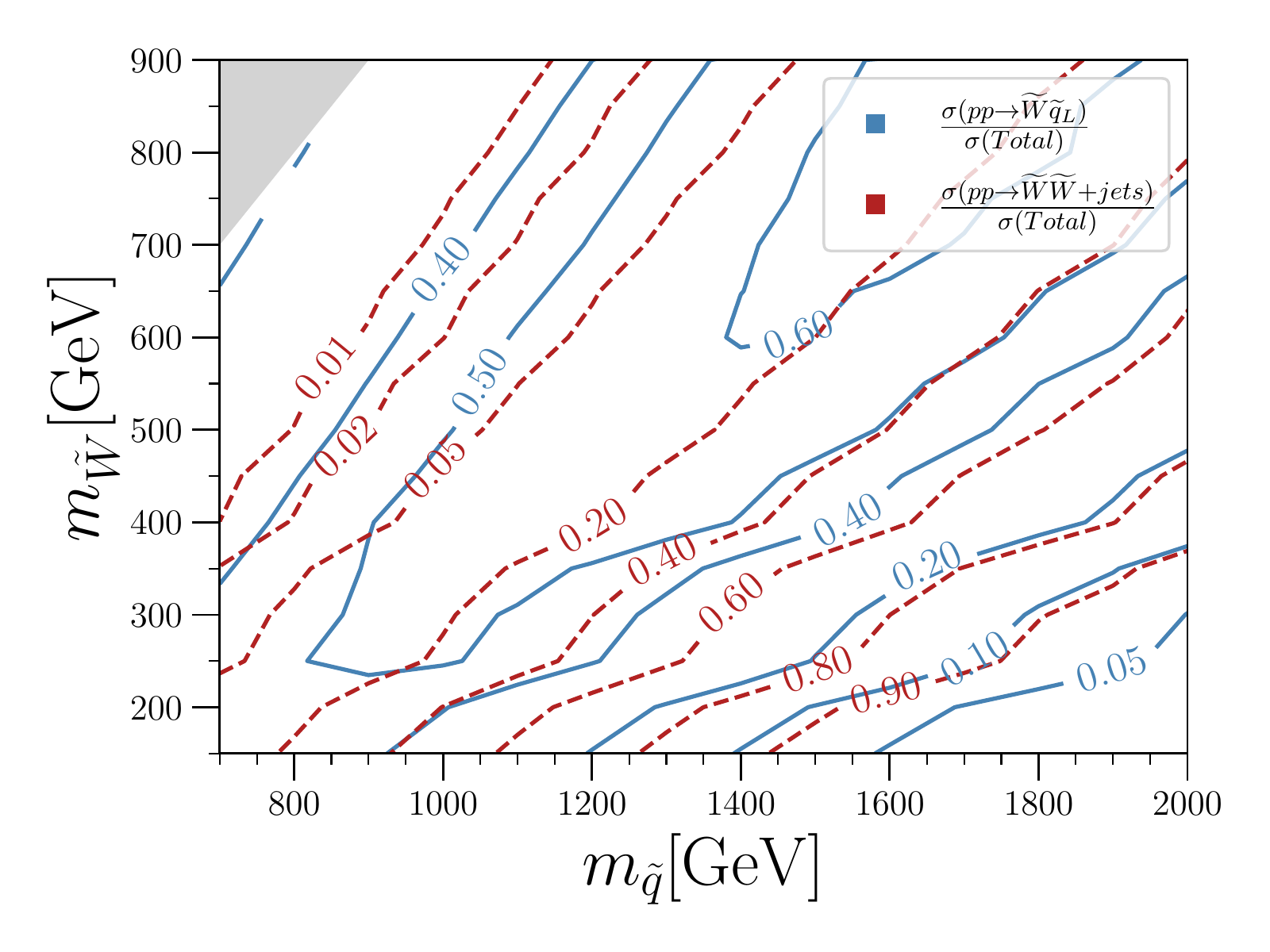}    
    \includegraphics[width=0.49\textwidth]{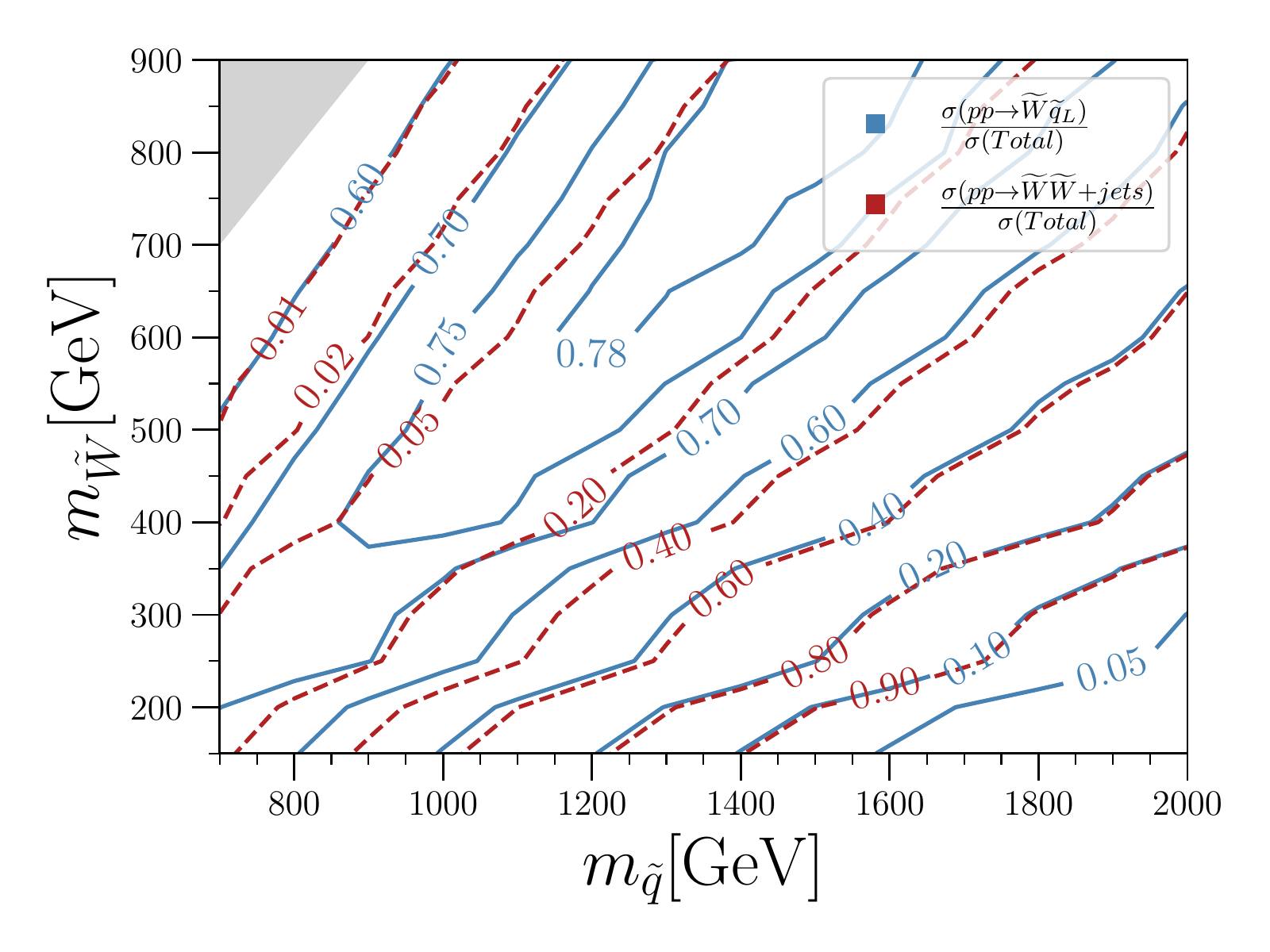}
    \caption{
    %The relative cross section of the $pp \to \tilde q_L \widetilde W $ channel with respect to the total cross section, which includes, $pp \to \tilde q_L \widetilde W, \tilde q \tilde q^*$ as well as $\widetilde W \widetilde W + {\jQCDs}$ final states.
    Relative production rates 
$\sigma(pp \to \wino \tilde q_L)/\sigma({\rm total})$ [blue-solid]
and 
$\sigma(pp \to \wino \wino + \jQCDs)/\sigma({\rm total})$ [red-dashed]
with $\sigma({\rm total}) = \sigma(pp \to \tilde q_L \tilde q_L,\, \wino \tilde q_L,\, \wino \wino + \jQCDs)$.
    The cross section of the last process is measured after imposing
    a cut $p_T(j^{\rm ISR}_1) > 200$ GeV on the hardest parton in the event generation. 
    The left and right plots correspond to
    the 8- and 2-flavour scenarios, respectively. }   \label{fig:xsec_QN_planes_total}
\end{figure}

Within our squark-electroweakino scenario, there is yet another production mode that contributes to the mono- and di-jet channels, $pp \to \tilde \chi \tilde \chi + \jQCDs$. For the bino-like LSP,  the production is induced by the $t$-channel squark exchange diagram and the cross section is negligible when the current squark mass limit is taken into account. The cross section is much larger for the wino- or higgsino-like LSP scenarios, since they are produced through Drell-Yan processes and multiple final states contribute.   For example, for the wino-like LSP case, three processes: $\tilde \chi \tilde \chi = \wino^+ \wino^-$, $\wino^0 \wino^+$ and $\wino^0 \wino^-$, contribute to this channel. %Figure~\ref{fig:xsec_QN_planes_total} displays the relative rate for the associated  squark-wino production channel with respect to the  total SUSY production channel, including $\tilde q_L \widetilde W$, $\tilde q \tilde q^*$ as well as $\widetilde W \widetilde W + \jQCDs$. The last process can contribute to the mono- and di-jet signal regions  only when the ISR jet is hard enough. We therefore impose a cut  $p_T(j^{\rm ISR}_1) > 200$\,GeV on the hardest parton in the event generation and estimate the cross section  after this cut. 
Fig.~\ref{fig:xsec_QN_planes_total} displays contours of the relative production rates 
$\sigma(pp \to \wino \tilde q_L)/\sigma({\rm total})$ [blue-solid]
and 
$\sigma(pp \to \wino \wino + \jQCDs)/\sigma({\rm total})$ [red-dashed]
with $\sigma({\rm total}) = \sigma(pp \to \tilde q_L \tilde q_L,\, \wino \tilde q_L,\, \wino \wino + \jQCDs)$.
The last process can contribute to the mono- and di-jet signal regions  only when the ISR jet is hard enough. We therefore impose a cut  $p_T(j^{\rm ISR}_1) > 200$\,GeV on the hardest parton in the event generation and estimate the cross section  after this cut. 
%Comparing Fig.~\ref{fig:xsec_QN_planes_total} with Fig.~\ref{fig:xsec_QN_planes_ratio}, we see that the $\widetilde W \widetilde W + {\jQCDs}$ cross section  is larger compared to the other processes, particularly in the region with the heavy squark and light wino. Nevertheless, around $(m_{\tilde q}, m_{\widetilde W}) = (1400, 600)$ GeV, the squark-wino associated production  is still the dominant channel and it takes up  more than 55 and 75 \% of the total cross section in the 8- and 2-flavour squark scenarios, respectively.  
We see that 
the $pp \to \widetilde W \widetilde W + {\jQCDs}$ process dominates ($\gtrsim 80 \, \%$) 
in the 
heavy-squark and light-wino region
(e.g.~$m_{\tilde q} \gtrsim 1.5$ TeV and $m_{\wino} \lesssim 300$ GeV).
On the other hand, around $(m_{\tilde q}, m_{\widetilde W}) = (1400, 600)$ GeV, the squark-wino associated production  is still the dominant channel and it takes up  more than 55 and 75 \% of the total cross section in the 8- and 2-flavour squark scenarios, respectively.

%%%%%%%%%%%%%%%%%%%%%%%%%%%%%%%%%%%%%%%%%
\subsection{Limit on the $(m_{\tilde q}, m_{\tilde \chi})$ mass plane}
\label{sec:limit}
%%%%%%%%%%%%%%%%%%%%%%%%%%%%%%%%%%%%%%%%%

To reveal the impact of the $\tilde q \widetilde W$ and $\widetilde W \widetilde W + \jQCDs$ productions on the squark searches at the LHC,
we recast ``an energetic jet\,+\,$E_T^{\rm miss}$'' \cite{ATLAS:2021kxv} and
``jets\,+\,$E_T^{\rm miss}$'' \cite{ATLAS:2020syg}
analyses of ATLAS, based on the ($\sqrt{s} = 13$ TeV, $L = 139$ fb$^{-1}$) data.
The former analysis targets several physics cases, including the dark matter direct production associated with ISR jets, $pp \to \chi_{_{\rm DM}} \chi_{_{\rm DM}} + \jQCDs$,
and the compressed squark-neutralino scenario with $(m_{\tilde q} - m_{\tilde \chi})/m_{\tilde q} \ll 1$.
The latter analysis are designed to look for gluinos and squarks in supersymmetric models and multiple signal regions are defined and used to cover a broad range of production and decay processes as well as mass assumptions.
Among them we find two signal regions particularly  
relevant to our scenario;
MB-C-2 and MB-SSd-2. 
The MB-C-2 (multi-bin compressed) signal region 
targets the compressed squark scenario as in 
``an energetic jet\,+\,$E_T^{\rm miss}$'' analysis.
The MB-SSd-2 (multi-bin squark-squark direct 2-jets)
signal region is designed to capture the signal from $pp \to \tilde q \tilde q^*$, followed by $\tilde q \to q \tilde \chi$.

\begin{table}[t!]
  \begin{minipage}[t]{.32\textwidth}
    \begin{center}
      \begin{tabular}{| c |} 
 \hline
 {\bf an energetic jet + $\met$} \\ 
 ($139$ fb$^{-1}$) \\
 \hline
 \hline
 $e, \mu, \tau, \gamma$ veto  \\ 
 \hline
 $\met > 200$  \\
 \hline
 $p_T^{j_1} > 150$, $|\eta^{j_1}| < 2.4$   \\
 ~   \\
 \hline
 $N_j(p_T > 30, |\eta|<2.8) \le 4$  \\
 \hline
 $\Delta \phi({\rm jet},{\bf p}_T^{\rm miss}) > 0.4$ (0.6)
 \\
 \hline
 $\met$ binned  \\
\hline
\end{tabular}
    \end{center}
  \end{minipage}
  \hfill
  \begin{minipage}[t]{.32\textwidth}
    \begin{center}
     \begin{tabular}{|c |} 
 \hline
{\bf jets + $\met$} ($139$ fb$^{-1}$) \\
{\bf MB-C-2} \\
 \hline\hline
 $e,\mu$ veto \\
 \hline
 $\met > 300$ \\
 \hline
 $p_T^{j_1} > 600$, $|\eta^{j_1}| < 2.8$ \\
% \hline
%  \\
 \hline
 $p_T^{j_2} > 50$, $|\eta^{j_2}| < 2.8$ \\
  \hline
 $N_j(p_T > 50, |\eta|<2.8) \le 3$  \\
 \hline
 %~\\
 $\Delta \phi({\rm jet},{\bf p}_T^{\rm miss}) > 0.4$  \\ 
 %~\\
 \hline
 ($\met/\sqrt{H_T}$, $m_{\rm eff}$) binned \\
 \hline
\end{tabular}
    \end{center}
  \end{minipage}
  \begin{minipage}[t]{.32\textwidth}
    \begin{center}
     \begin{tabular}{|c |} 
 \hline
{\bf jets + $\met$} ($139$ fb$^{-1}$) \\
{\bf MB-SSd-2} \\
 \hline\hline
 $e,\mu$ veto \\
 \hline
 $\met > 300$ \\
 \hline
 $p_T^{j_1} > 250$, $|\eta^{j_1}| < 2$ \\
 \hline
 $p_T^{j_2} > 250$, $|\eta^{j_2}| < 2$ \\
% \hline
%  \\
 \hline
 $N_j(p_T > 50, |\eta|<2.8) \le 3$  \\
 \hline
 %~\\
 $\Delta \phi({\rm jet},{\bf p}_T^{\rm miss}) > 0.8$  \\ 
 %~\\
 \hline
 ($\met/\sqrt{H_T}$, $m_{\rm eff}$) binned \\
 \hline
\end{tabular}
    \end{center}
  \end{minipage}
\caption{Event selection criteria in ``an energetic jet + $\met$'' (left) 
and the MB-C-2 (middle) and MB-SSd-2 (right) signal regions in ``jets + $\met$'' analyses 
\cite{ATLAS:2021kxv, ATLAS:2020syg}, respectively.
In the left table the threshold of $\Delta \phi$ cut
is raised to 0.6 if $\met < 250$GeV.
The unit of energy and momentum is GeV.
}
\label{tb:cut}  
\end{table}

The event selection criteria of these three signal regions are summarised in Table \ref{tb:cut}.
There are commonalities among them.
In all signal regions,  events with isolated leptons are vetoed.
Large $\met$ ($> 200$ or 300 GeV) 
and a couple of high $p_T$ jets are required.
Those high $p_T$ jets must have a large angular separation, $\Delta \phi$, in the transverse plane from the direction of missing momentum, ${\bf p}_T^{\rm miss}$.
The ``an energetic jet + $\met$'' analysis
and the MB-C-2 signal region in the ``jets + $\met$'' analysis
are monojet-type and require one particularly energetic jet
compared to the other jets.
MB-SSd-2 is, on the other hand, dijet-type since it demands 
two very energetic jets.
All three signal regions limit the number of jets:
the ``an energetic jet + $\met$'' analysis allows up to four jets 
with $p_T > 30$ GeV and $|\eta| < 2.8$,
while MB-C-2 and MB-SSd-2 accept only up to three jets with  
$p_T > 50$ GeV and $|\eta| < 2.8$.
In the ``an energetic jet + $\met$'' analysis,
after imposing the conditions listed in Table \ref{tb:cut},
$\met$/GeV is sliced into the bins with upper and lower thresholds [200, 250, 300, 350, 400, 500, 600, 700, 800, 900, 1000, 1100, 1200, $\infty$] and data is analysed 
exclusively and inclusively with these bins.
In the ``jets + $\met$'' analysis 
the signal events are analysed with the 2-dimensional bins:
$\met/\sqrt{H_T} = [16, 22, \infty]\,{\rm GeV}^{1/2}$
and
$m_{\rm eff}/{\rm TeV} = [1.6, 2.2, 2.8, \infty]$
for the MB-C-2 signal region,
and
$\met/\sqrt{H_T} = [10, 16, 22, 26, \infty]\,{\rm GeV}^{1/2}$
and
$m_{\rm eff}/{\rm TeV} = [1.0, 1.6, 2.2, 2.8, 3.4, 4.0, \infty]$
for the MB-SSd-2 signal region,
where $H_T$ is defined as
the scalar sum of transverse momenta
of all jets with $p_T > 50$ GeV and $|\eta| < 2.8$
and $m_{\rm eff} \equiv H_T + \met$.

The current best exclusion limit in the squark-neutralino mass plane 
is obtained by interpreting the results 
in these three signal regions for 
the $pp \to \tilde q \tilde q^* \to (q \tilde \chi) (q \tilde \chi)$ process, 
assuming 
the neutralino is bino-like and ignoring the 
$pp \to \tilde q \tilde \chi$ and
$pp \to \tilde \chi \tilde \chi + \jQCDs$ channels.
Our goal is to estimate the impact of the
$pp \to \tilde q \widetilde W$ and
$pp \to \widetilde W \widetilde W + \jQCDs$ processes
on the mass limit 
in the squark-wino scenario.
In our simulation, we find the sensitivity 
of the MB-C-2 and MB-SSd-2 signal regions in the ``jets + $\met$'' 
analysis 
are always higher than
that of the 
``an energetic jet + $\met$'' analysis.
In what follows, we therefore concentrate on the former analysis.

\begin{figure}[t!]
    \centering
    \includegraphics[width=0.49\textwidth]{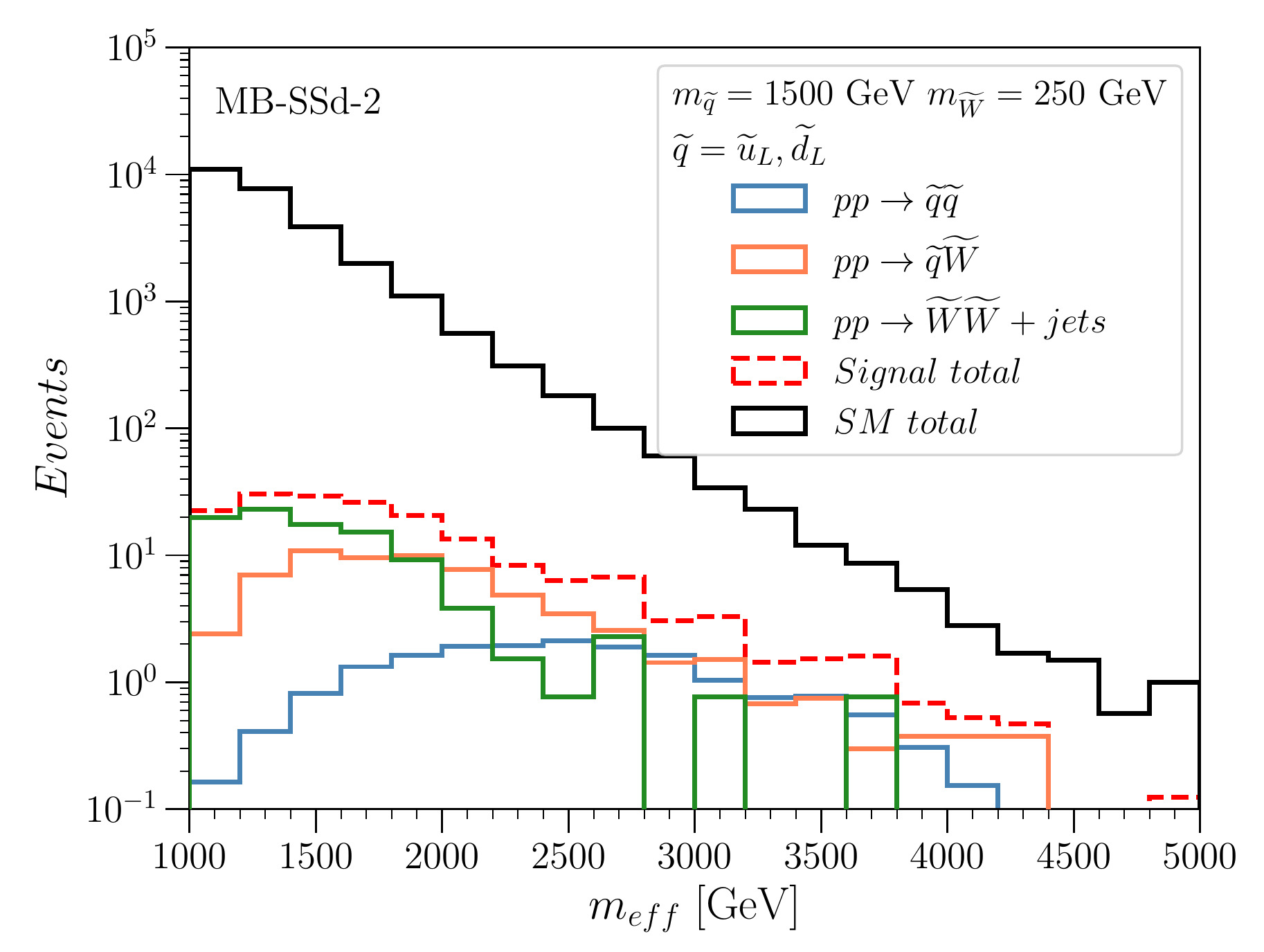}
    \includegraphics[width=0.49\textwidth]{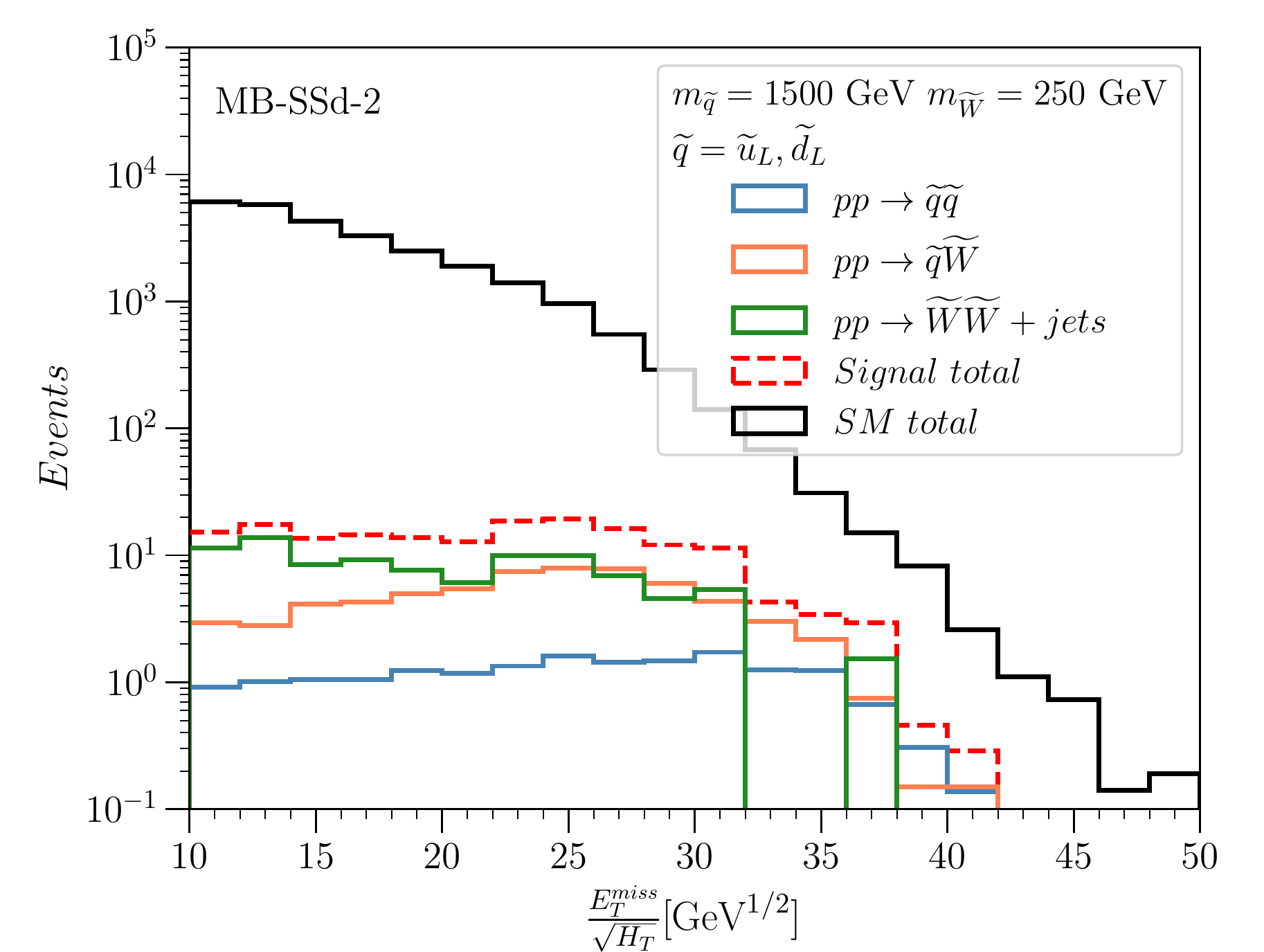}
    \includegraphics[width=0.49\textwidth]{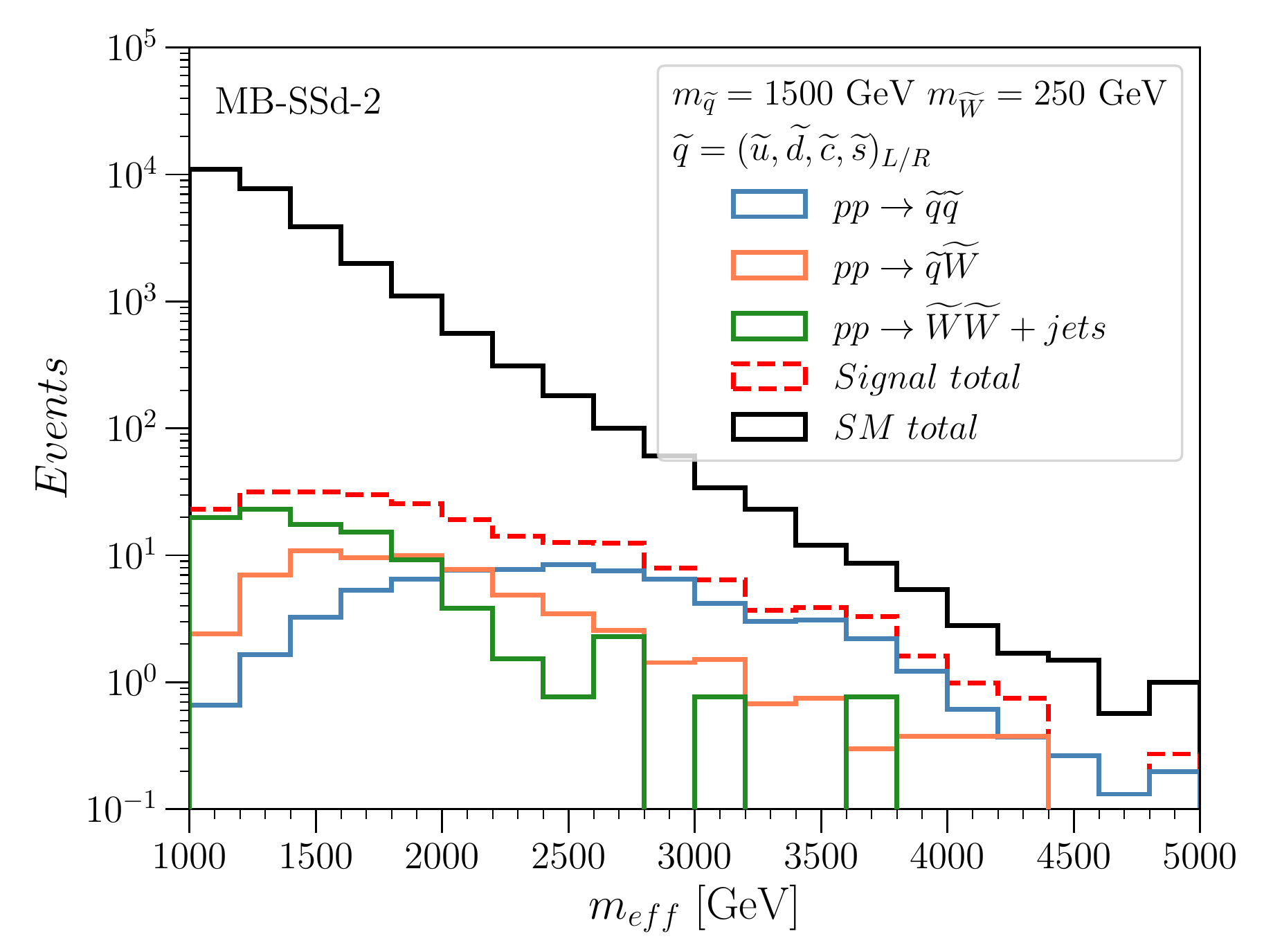}
    \includegraphics[width=0.49\textwidth]{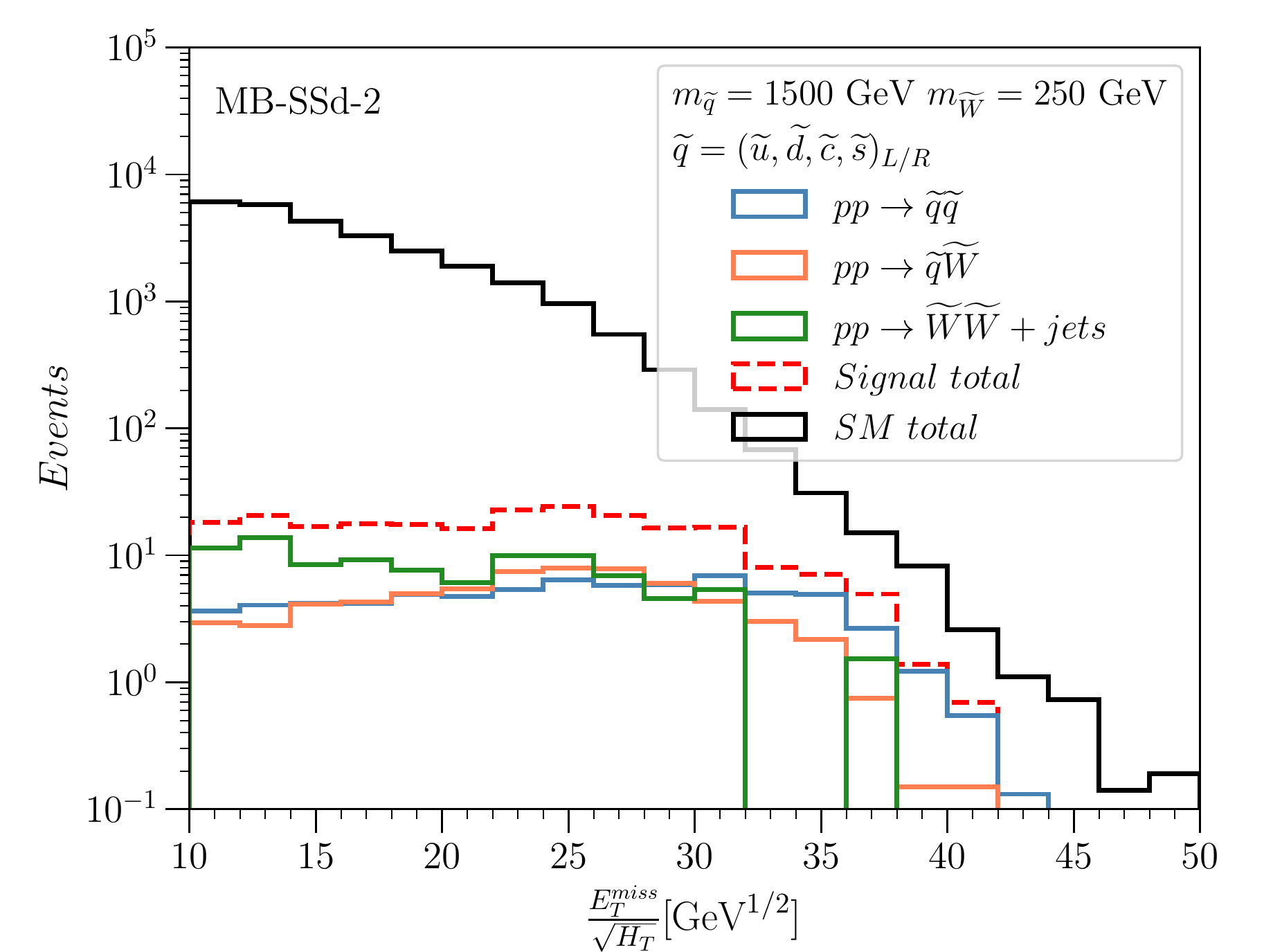}
    \caption{The $m_{\rm eff}$ (left) and $\met/\sqrt{H_T}$ (right) distributions
    in the MB-SSd-2 signal region in the 2-flavour (top) and 8-flavour (bottom) scenarios.}
    \label{fig:QW_kinemtic}
\end{figure}

In Fig.~\ref{fig:QW_kinemtic} we show 
the $m_{\rm eff}$ (left) and $\met/\sqrt{H_T}$ (right)
distributions from the Standard Model (black),
$pp \to \tilde q \tilde q^*$ (blue),
$pp \to \tilde q \widetilde W$ (orange),
and
$pp \to \widetilde W \widetilde W + \jQCDs$ (green)
production channels in the 2-flavour (top) and 8-flavour (bottom)
scenarios. The squark and wino masses are fixed at 1500 and 250 GeV, respectively.  
The distributions are obtained after imposing the event selection  
in the MB-SSd-2 signal region except for the requirement on 
$\met/\sqrt{H_T}$ and $m_{\rm eff}$.
The histograms for the Standard Model background are taken from Ref.~\cite{ATLAS:2020syg}.
The distributions of SUSY processes are 
obtained by the following simulation pipe-line:
signal events are generated at parton-level using {\tt MadGraph5\_aMC@NLO 3.1.0}\cite{madgraph5} with up to two additional partons in the final state and with the NNPDF23LO~\cite{Ball:2012cx,Buckley:2014ana} PDF set. The event samples are passed to {\tt Pythia-8.244} \cite{Sjostrand:2014zea} to simulate decays of SUSY particles, the parton shower and hadronisation. Jet matching and merging to parton-shower calculations is accomplished by the MLM algorithm~\cite{Mangano:2006rw}. The detector simulation and jet clustering~\cite{Cacciari:2011ma,Cacciari:2005hq} are performed with {\tt Delphes 3} \cite{deFavereau:2013fsa}
within {\tt CheckMATE 2} \cite{checkmate1,checkmate2,analysismanager}. 

We see from all plots in Fig.~\ref{fig:QW_kinemtic} that
there is a tendency 
that
the $\tilde q \wino$
and 
$\wino \wino + \jQCDs$
production modes 
dominate
at lower values of $m_{\rm eff}$ and $\met/\sqrt{H_T}$.
On the other hand, 
at higher values
the contribution from the $\wino \wino + \jQCDs$ mode is suppressed, while
the $\tilde q \tilde q^*$ mode
becomes important especially for the 8-flavour case.
In the $m_{\rm eff}$ distributions,
the $\tilde q \wino$ and $\tilde q \tilde q^*$ modes dominate 
in the $m_{\rm eff} \gtrsim 3$ (2.5) TeV region for the 2 (8) flavour case.
The distributions have peaks at $\sim 2.5$ TeV 
for $\tilde q \tilde q^*$
and $\sim 1.5$ TeV for $\tilde q \widetilde W$ channels, 
while peaks are not visible for the $\wino \wino + \jQCDs$ channel.
In the $\met/\sqrt{H_T}$ distributions,
all three signal sub-processes are comparable in the $\met/\sqrt{H_T} \gtrsim 30$ region.

\begin{figure}[t!]
    \centering
    \includegraphics[width=0.49\textwidth]{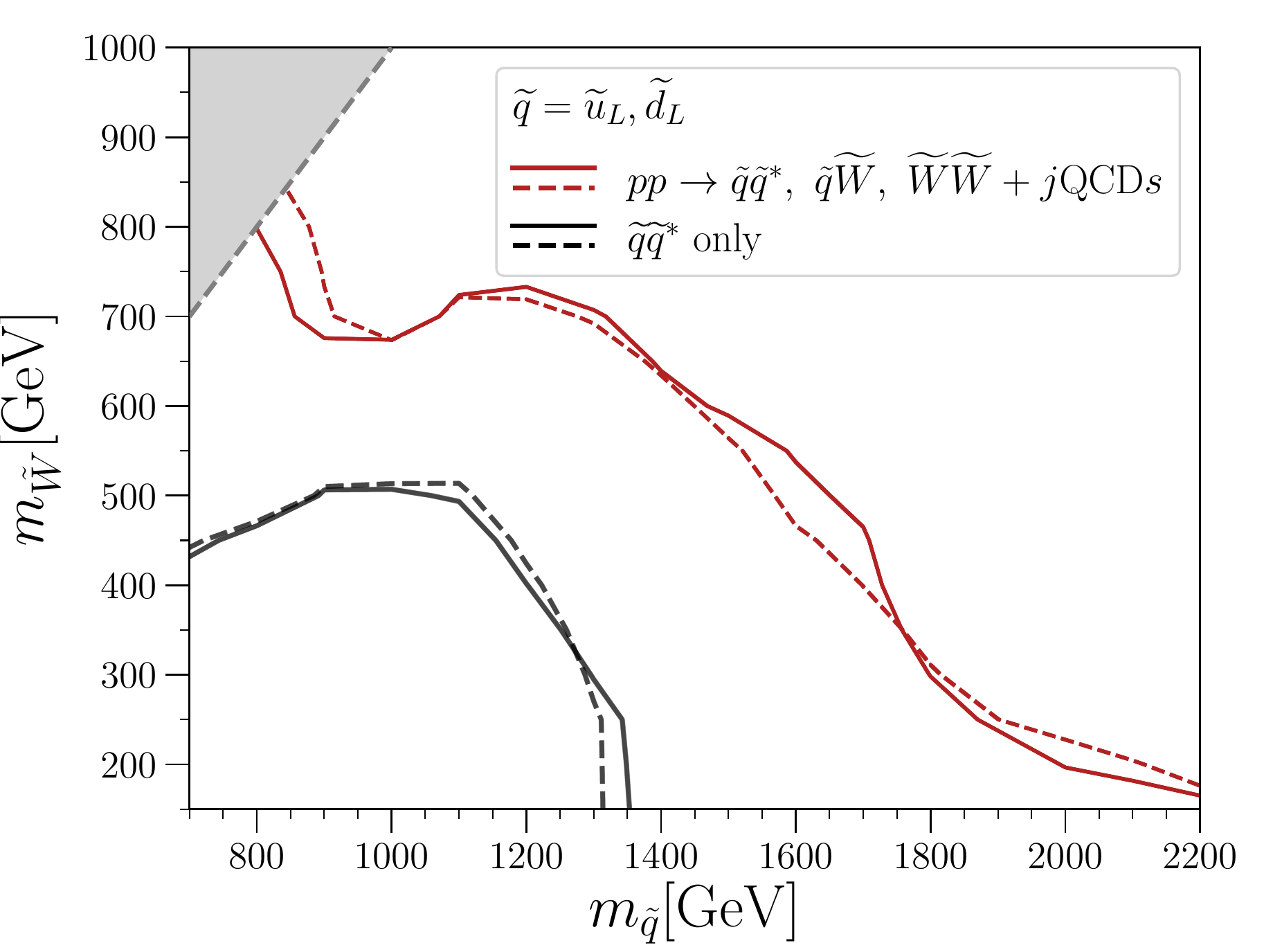}
    \includegraphics[width=0.49\textwidth]{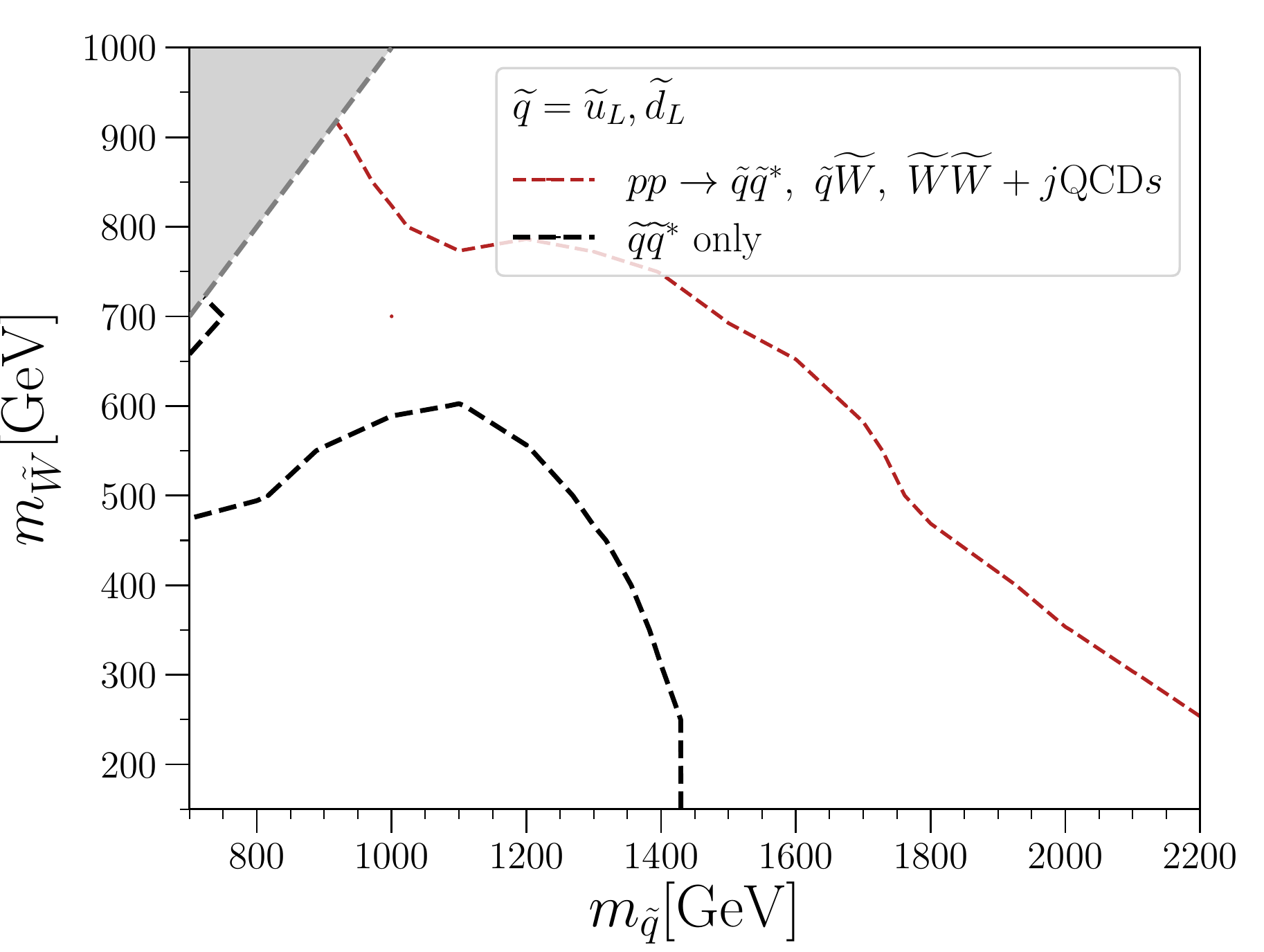}
    \includegraphics[width=0.49\textwidth]{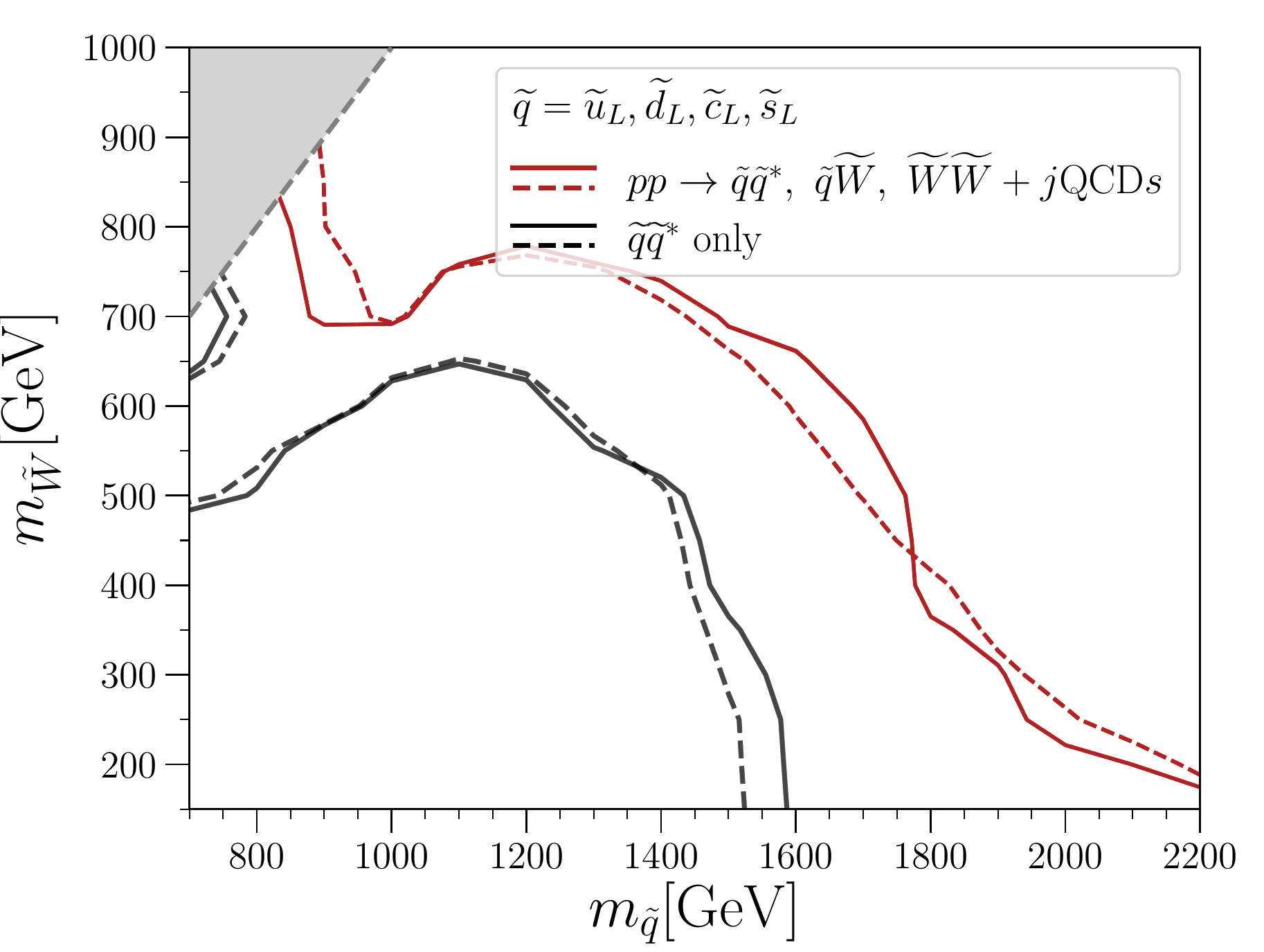}
    \includegraphics[width=0.49\textwidth]{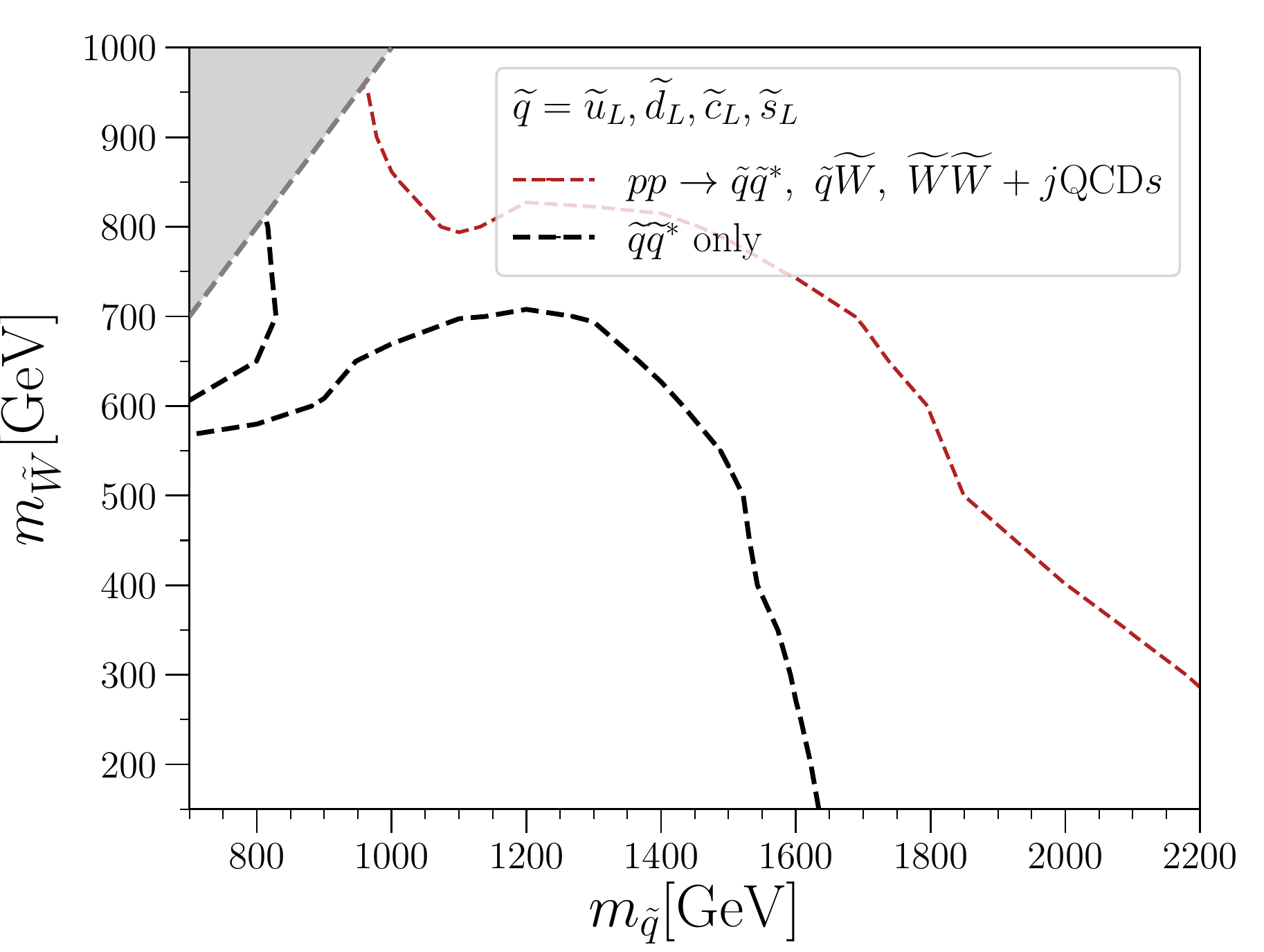}
    \includegraphics[width=0.49\textwidth]{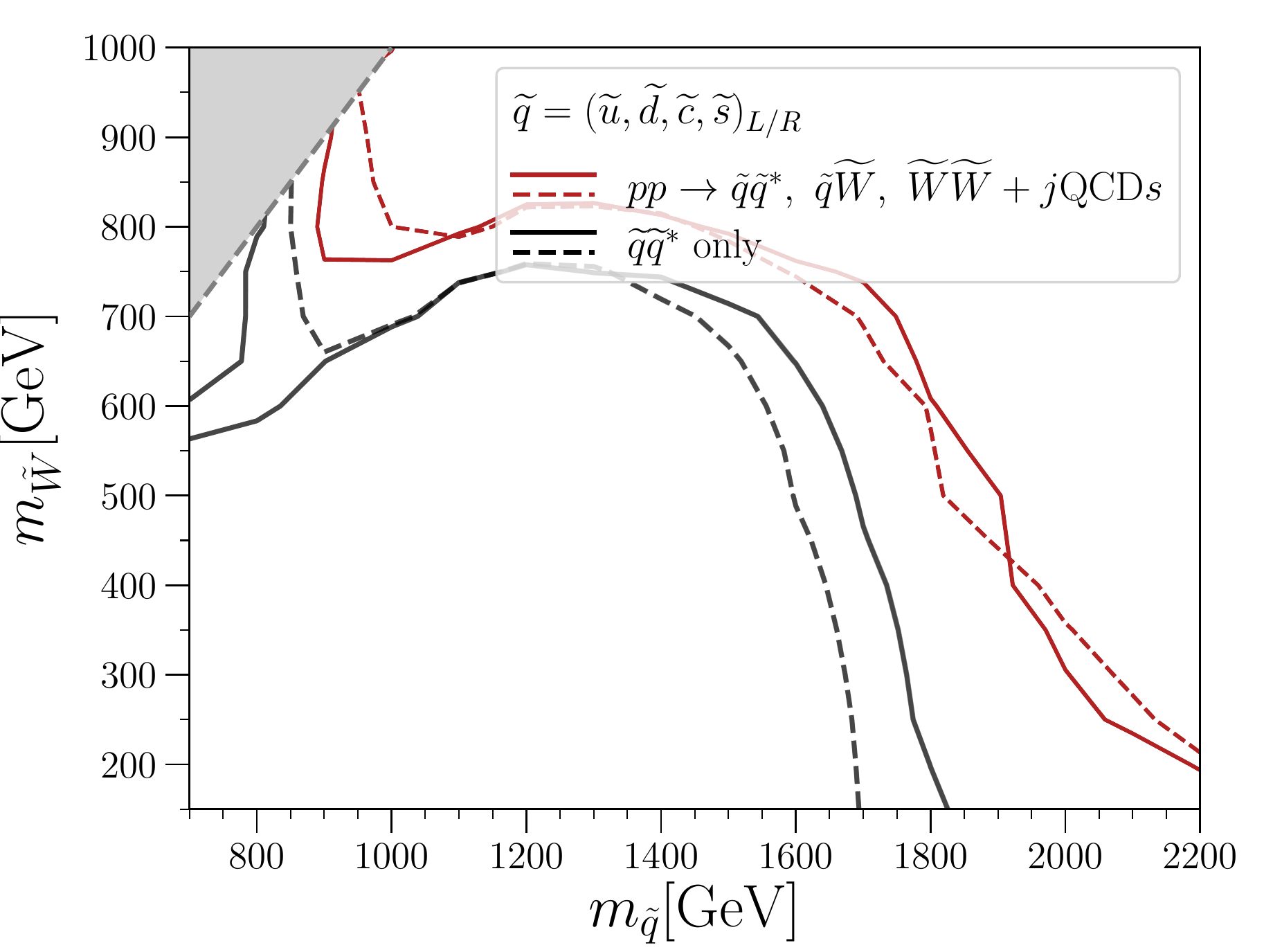}
    \includegraphics[width=0.49\textwidth]{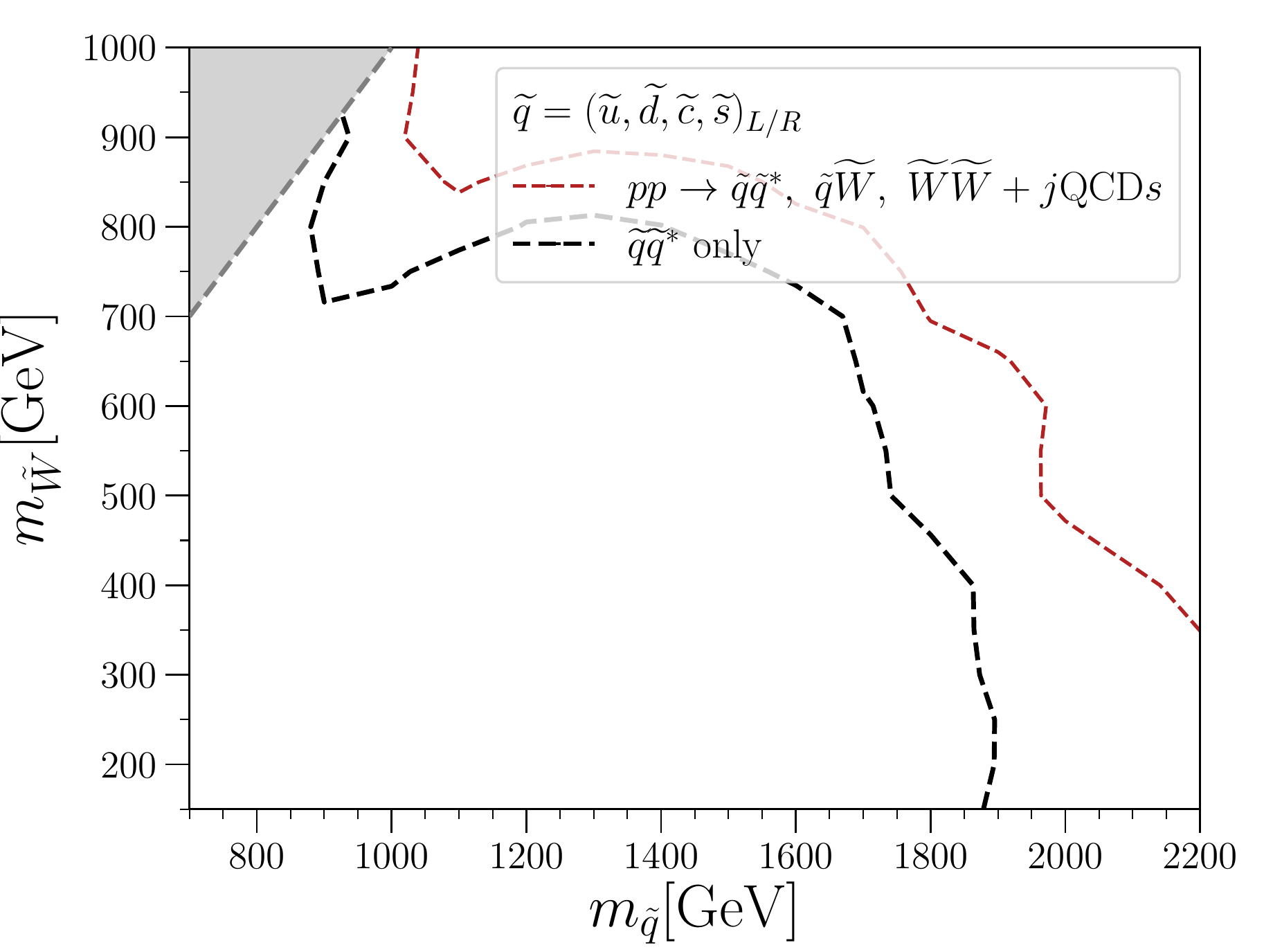}
    \caption{Left; the 95\% CL exclusion limit on the $(m_{\tilde q}, m_{\widetilde W})$ plane obtained by recasting the ATLAS ``jets + $\met$'' analysis \cite{ATLAS:2020syg} with 139 fb$^{-1}$ data.
    The solid curves represent the limit calculated with the observed data, while the dashed curves correspond to the expected limit assuming the observed data exactly coincides with the Standard Model expectation. The black curves represent the limit obtained if only the squark pair production is considered,
    while the red curves correspond to the limit including all production channels: $pp \to \tilde q \tilde q^*, \tilde q \widetilde W$ and $\widetilde W \widetilde W + \jQCDs$.
    The top, middle and bottom panels correspond to 
    the 2, 4 and 8 flavour squark scenarios, respectively.
    Right; the 95\% CL projected sensitivity on the $(m_{\tilde q}, m_{\widetilde W})$ plane obtained by rescaling the ATLAS ``jets + $\met$'' analysis \cite{ATLAS:2020syg} to the Run-3 with 300 fb$^{-1}$ (13 TeV).   }
    \label{fig:sq_W_lim}
\end{figure}

In the left panels of Fig.~\ref{fig:sq_W_lim}
we show the observed (solid curve)
and expected (dashed curve) 
95\% CL exclusion contours
on the ($m_{\tilde q}, m_{\widetilde W}$)
plane obtained using {\tt CheckMATE}
with the data corresponding to the full Run-2 (139 fb$^{-1}$) integrated luminosity.
In the calculation we have included 
the ATLAS ``an energetic jet + $\met$'' 
analysis \cite{ATLAS:2021kxv} as well as 
all cut-and-count based signal regions
defined in the ATLAS ``jets + $\met$''
analysis \cite{ATLAS:2020syg}.
We find, however, the strongest constraint comes 
from the MB-SSd-2 and MB-C-2 signal regions across the plane.
In order to maximise
the sensitivity, we performed multi-binned analysis within the MB-SSd-2 and MB-C-2 signal regions
using the {\tt HistFitter-2.0} package~\cite{Baak:2014wma}.
For each orthogonal bin defined in the signal region $\alpha$, 
we use the following four quantities:
(1) the number of SM background and (2) the number of observed events 
are taken from the ATLAS paper \cite{ATLAS:2020syg},
while
(3) the number of signal events and (4) the corresponding uncertainty are estimated using the simulation pipeline mentioned above. 

The profile likelihood is constructed using
these quantities ($4 \times $[the number of bins]) combining the bins.
A test is performed to evaluate the $p$-value as a function of the signal strength, $\mu$,
an overall normalisation of the signal events.
This allows to determine the 95\% confidence level (CL) limit on the signal strength, $\mu_{95}^{\rm obs}(\alpha)$, for signal region $\alpha$.
In a similar way, the {\it expected} limit, $\mu_{95}^{\rm exp}(\alpha)$, is obtained by substituting the SM background for the observed events in the above procedure. 
The \emph{best expected} (or the most sensitive) signal region, $\alpha^*$, for the mass point is defined 
as the signal region that gives the smallest $\mu_{95}^{\rm exp}(\alpha)$, i.e.\ 
$\forall \alpha;~ \mu_{95}^{\rm exp}(\alpha^*) \le \mu_{95}^{\rm exp}(\alpha)$. The mass point is excluded at 95\% CL if $\mu_{95}^{\rm obs}(\alpha^*) < 1$.

The top, middle and bottom
panels in Fig.~\ref{fig:sq_W_lim}
correspond to the
2, 4 and 8 flavour squark scenarios, respectively.
The black curves represent 
the exclusion limits, which are obtained by considering only the squark pair production channel.\footnote{We have validated that the black curve in the 8-flavour case agrees with the corresponding exclusion plot in the ATLAS analysis \cite{ATLAS:2020syg} in the high mass limit. In the compressed region at the squark mass of 800 GeV, our limit appears to be somehat weaker, which may be due to differences in the signal modelling.}
Those exclusion contours  
should be compared with 
the red contours, 
which represent the exclusion limits,
including all production channels, 
$\tilde q \tilde q^*$, 
$\tilde q \widetilde W$ and $\widetilde W \widetilde W + \jQCDs$,
in the wino-like LSP scenario. 
We see that the exclusion limit 
is extended both in the directions 
of $m_{\tilde q}$ and $m_{\widetilde W}$.
The limit is improved particularly
in the light wino heavy squark region,
in which the relative contributions from the $\tilde q \tilde W$
and $\tilde W \tilde W + \jQCDs$
are most enhanced.
In particular, the contribution from
the $\tilde W \tilde W + \jQCDs$
process 
is independent of the squark mass
and the region with $m_{\widetilde W} \lesssim 150$ GeV is excluded by this process alone regardless of the squark mass.
In the compressed mass region ($m_{\tilde q} \simeq m_{\widetilde W}$),
the impact of the $\tilde q \widetilde W$ and $\widetilde W \widetilde W + \jQCDs$ channels
is again large.
In this region, squark decays only produce soft particles and QCD radiation is the unique source of high $p_T$ jets.
This makes the efficiencies of 
the three production channels almost the same.
In the intermediate region 
with $m_{\tilde q} \sim 2 m_{\widetilde W}$, the impact of the 
$\tilde q \widetilde W$ and $\widetilde W \widetilde W + \jQCDs$ processes 
is modest.
In this region, high $p_T$ jets are obtained from the squark decay
and the acceptance of the $\tilde q \tilde q^*$ channel becomes the largest.
However, the gain in the exclusion limit is still significant due to the contribution from the $\tilde q \widetilde W$ channel, 
of which
production rate and acceptance are still comparable with those of the $\tilde q \tilde q^*$ channel.

As discussed earlier,
the relative contribution of the $\tilde q \widetilde W$ 
and $\widetilde W \widetilde W + \jQCDs$
processes with respect of $\tilde q \tilde q^*$ is larger for the 4-flavour case (where all left-type squarks are decoupled) than the 8-flavour.
The relative contribution is even more enhanced in the 2-flavour scenario,
where all squarks except for the first generation left-type doublet ($\tilde u_L$ and $\tilde d_L$) are decoupled.  
We see in the left panels of Fig.~\ref{fig:sq_W_lim}
that by including the $\tilde q \widetilde W$ and $\widetilde W \widetilde W + \jQCDs$ channels the squark mass limit is extended by $\sim 45$, 20 and 10\,\% 
for the 2, 4 and 8-flavour squark scenarios, respectively,
at $m_{\widetilde W} \simeq 400$ GeV.

The limits combining all three signal sub-processes (red) are significantly extended 
compared to the $\tilde q \tilde q^*$-only limits (black)
in the compressed mass region, $m_{\tilde q} \sim m_{\wino}$.
We checked that MB-C-2 signal region is sensitive to the compressed mass region, while the constraint from MB-SSd-2 is sensitive to the region with larger $\Delta m = m_{\tilde q} - m_{\wino}$.
In this region the efficiency of $\tilde q \tilde q^*$ is similar to that of 
$\tilde q \wino$ and $\wino \wino + \jQCDs$ since squark decays do not produce high $p_T$ jets and
the relative importance of the latter two processes is therefore enhanced in this region. 
One interesting consequence of this is
that one can find the {\it wino mass independent} lower bounds on the squark mass, $m_{\tilde q} \gtrsim 800$, 850 and 900 GeV for the 2-, 4- and 8-flavour cases, respectively, once the
$\tilde q \wino$ and $\wino \wino + \jQCDs$ processes are included.
The limit on the squark mass is outside the plot range when the associate squark-wino production is omitted.

Finally in the right panels of Fig.~\ref{fig:sq_W_lim} we show the projected 95\% CL limits on the $(m_{\tilde q}, m_{\widetilde W})$ plane expected at Run-3 LHC with the integrated luminosity of 300 fb$^{-1}$. 
For this projection we take the conservative assumption that both signal and backgrounds grow linearly with the luminosity.
We see that the expected limit extends only mildly from Run-2 (139 fb$^{-1}$) to Run-3 (300 fb$^{-1}$). However, the impact of the $\tilde q \widetilde W$ and $\widetilde W \widetilde W + \jQCDs$ channels  is significantly larger at Run-3 in the high $m_{\tilde q}$ low $m_{\widetilde W}$ region. 
Compared to the $\tilde q \tilde q^*$ only scenario,  the mass reach is increased  by $\sim 45$, 30 and 15\,\% for the 2-, 4- and 8-flavour squark scenarios, respectively, at $m_{\widetilde W} \simeq 400$ GeV.

%%%%%%%%%%%%%%%%%%%%%%%%%%%%%%%%%%%%%%%%%
\section{Monojet from gluino-squark productions}
\label{sec:gl-sq}
%%%%%%%%%%%%%%%%%%%%%%%%%%%%%%%%%%%%%%%%%

\begin{figure}[t!]
\centering
\subfloat[]{
      \includegraphics[width=0.4\textwidth]{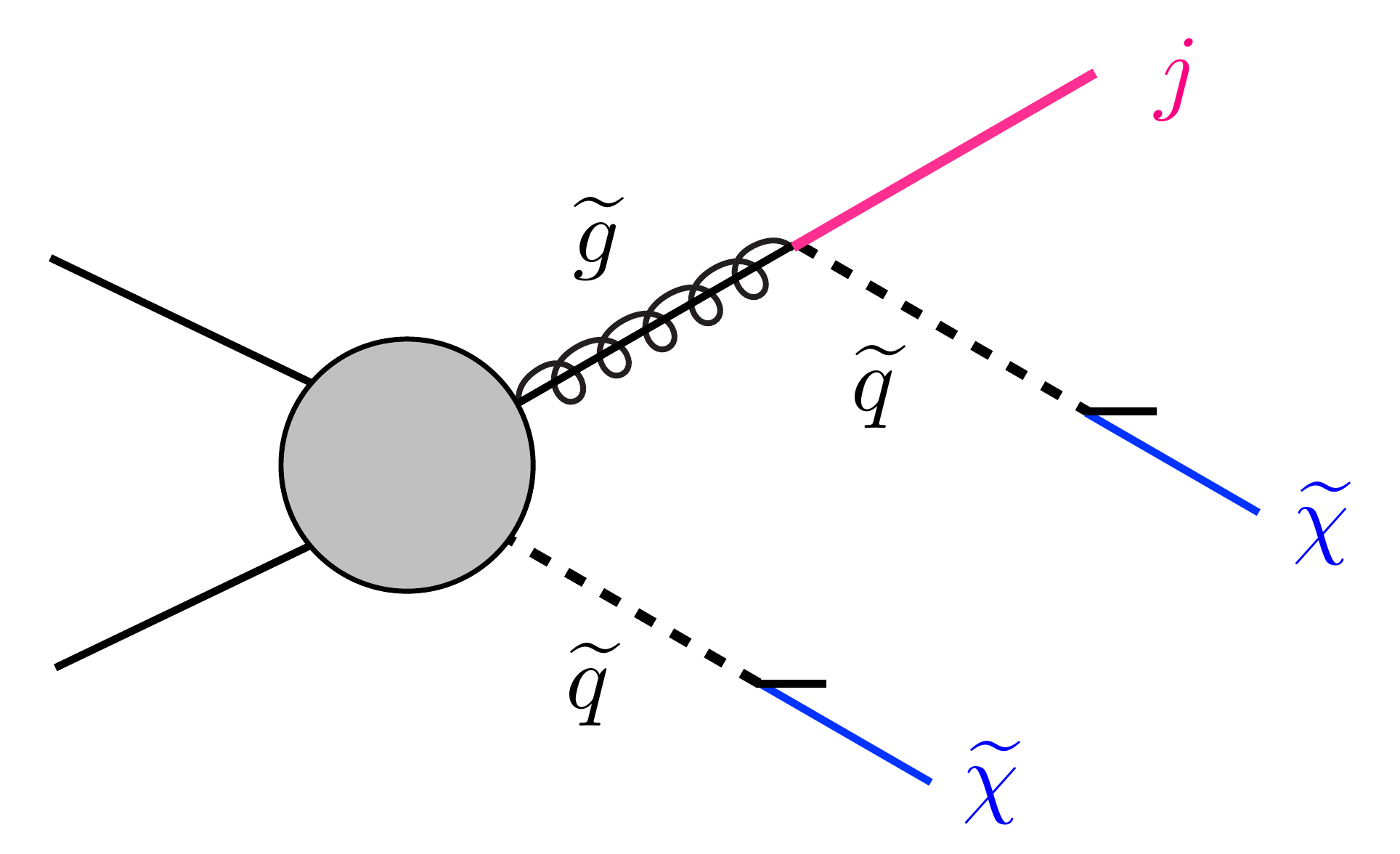}
\label{fig:monojet_col_a}
     }
\hspace{10mm}
\subfloat[]{
      \includegraphics[width=0.4\textwidth]{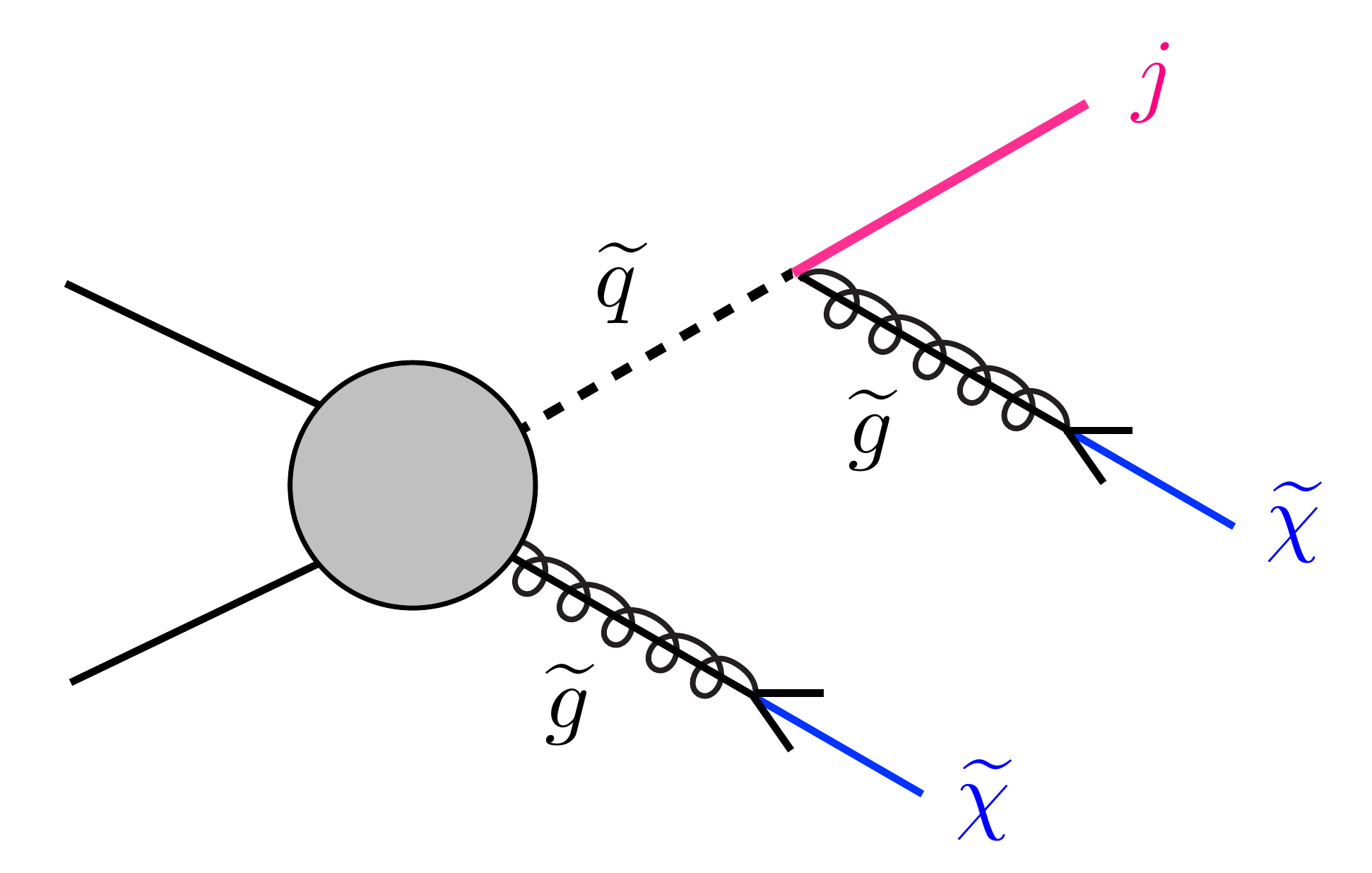}
\label{fig:monojet_col_b}
     }
\hfill
\caption{\label{fig:monojet_col}
Monojet signature from gluino-squark associated productions, where
the lighter of gluino and squark is only slightly heavier than
the LSP neutralino.
{\bf (a)} $m_{\tilde g} > m_{\tilde q} \simeq m_{\tilde \chi_1^0}$ {\bf (b)} $m_{\tilde q} > m_{\tilde g} \simeq m_{\tilde \chi_1^0}$.
}
\end{figure}

In this section, we investigate the 
constraint from the mono- and di-jet channels 
on the gluinos and squarks production.
As mentioned in the previous section, 
a distinctive monojet-like signature may be obtained
from these production modes
if the lightest neutralino is almost mass-degenerate with the lighter of the gluino and squark,
$m_{\tilde \chi_1^0} \simeq {\rm min}(m_{\tilde g}, m_{\tilde q})$. 
Those quasi-mass-degenerate spectra are required in the gluino-bino and squark-bino coannihilation scenarios \cite{DeSimone:2014qkh,Profumo:2004wk,Harigaya:2014dwa,Ellis:2015vaa}.
We consider 
the following three processes depending on the gluino-squark mass hierarchy:
\begin{center}
\renewcommand{\arraystretch}{1.2}
\begin{tabular}{ l l c l l}
Case $m_{\tilde g} > m_{\tilde q}$\,: & &~~~~~~~~~~& Case $m_{\tilde q} > m_{\tilde g}$\,: & \\ 
$pp \to \tilde g \tilde q$,~ $\tilde g \to q \tilde q$ & $\cdots$ ({\bf a1}) && 
$pp \to \tilde g \tilde q$,~ $\tilde q \to q \tilde g$ & $\cdots$ ({\bf b1}) \\  
$pp \to \tilde g \tilde g$,~ $\tilde g \tilde g \to (q \tilde q)(q \tilde q)$ & $\cdots$ ({\bf a2})&& 
$pp \to \tilde q \tilde q$,~ $\tilde q \tilde q \to (q \tilde g)(q \tilde g)$ & $\cdots$ ({\bf b2}) \\   
$pp \to \tilde q \tilde q + \jQCDs$ & $\cdots$ ({\bf a3})&& 
$pp \to \tilde g \tilde g + \jQCDs$ & $\cdots$ ({\bf b3})
\end{tabular}
\end{center}
We specifically mention ISR for ({\bf a3}) and ({\bf b3}) since it crucially contributes to the visible final state for these processes. At the event generation level, however, all processes include additional radiation and the same procedure is employed as explained in Section~\ref{sec:limit}. 

All of these processes potentially contribute to the mono- and di-jet channels.
The processes ({\bf a1}) and ({\bf b1}) correspond to
the associated squark-gluino production, followed by the decay of the heavier coloured particle to the lighter one, as discussed in the Introduction. 
These processes are also illustrated in Fig.~\ref{fig:monojet_col_a} and \ref{fig:monojet_col_b}, respectively. 
In this scenario, gluinos are generally not decoupled and therefore 
both squark-squark and squark-antisquark productions are important.
We simply write $pp \to \tilde q \tilde q$
to denote the inclusive process containing both squark-squark
and squark-antisquark productions throughout this section.
We also assume
all 8 flavour squarks
($\tilde u_L$, $\tilde d_L$, $\tilde s_L$, $\tilde c_L$, $\tilde u_R$, $\tilde d_R$, $\tilde s_R$, $\tilde c_R$)
have an equal mass, $m_{\tilde q}$.
It should be noted that
the processes ({\bf a3}) and ({\bf b3})
can contribute to the relevant signal regions only if the $\jQCDs$ is hard enough.
We therefore demand $p_T(j_1^{\rm ISR}) > 200$ GeV at parton level in the event generation for the ({\bf a3}) and ({\bf b3}) processes.
Although not explicitly denoted, the other processes, 
({\bf a1}), ({\bf a2}), ({\bf b1}) and ({\bf b2}), are also generated accompanied with the initial state QCD radiation.
However, no explicit cut is imposed on the ISR jets for these processes.

\begin{figure}[t!]
\centering
      \includegraphics[width=0.48\textwidth]{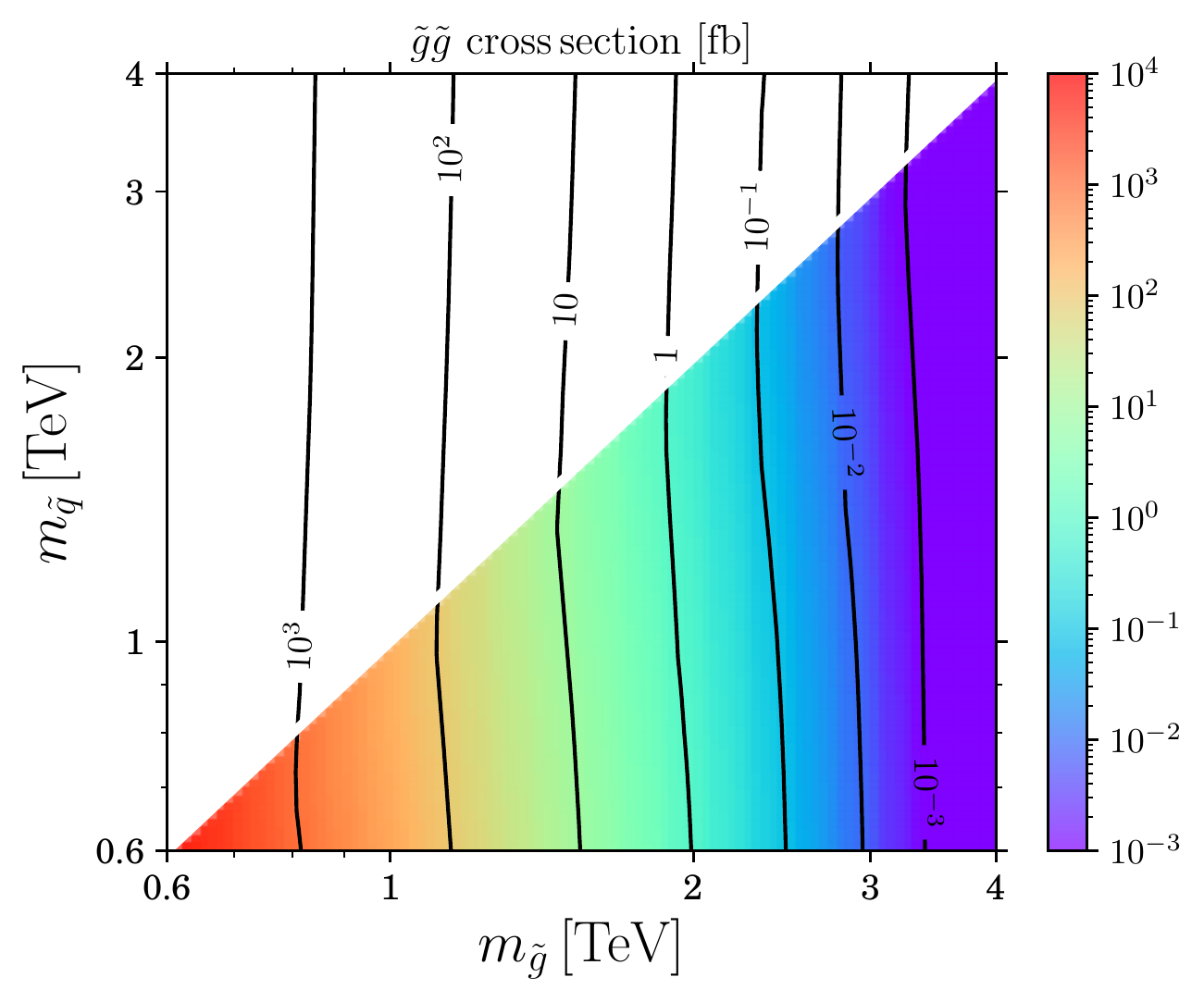}
\hspace{1mm}
      \includegraphics[width=0.48\textwidth]{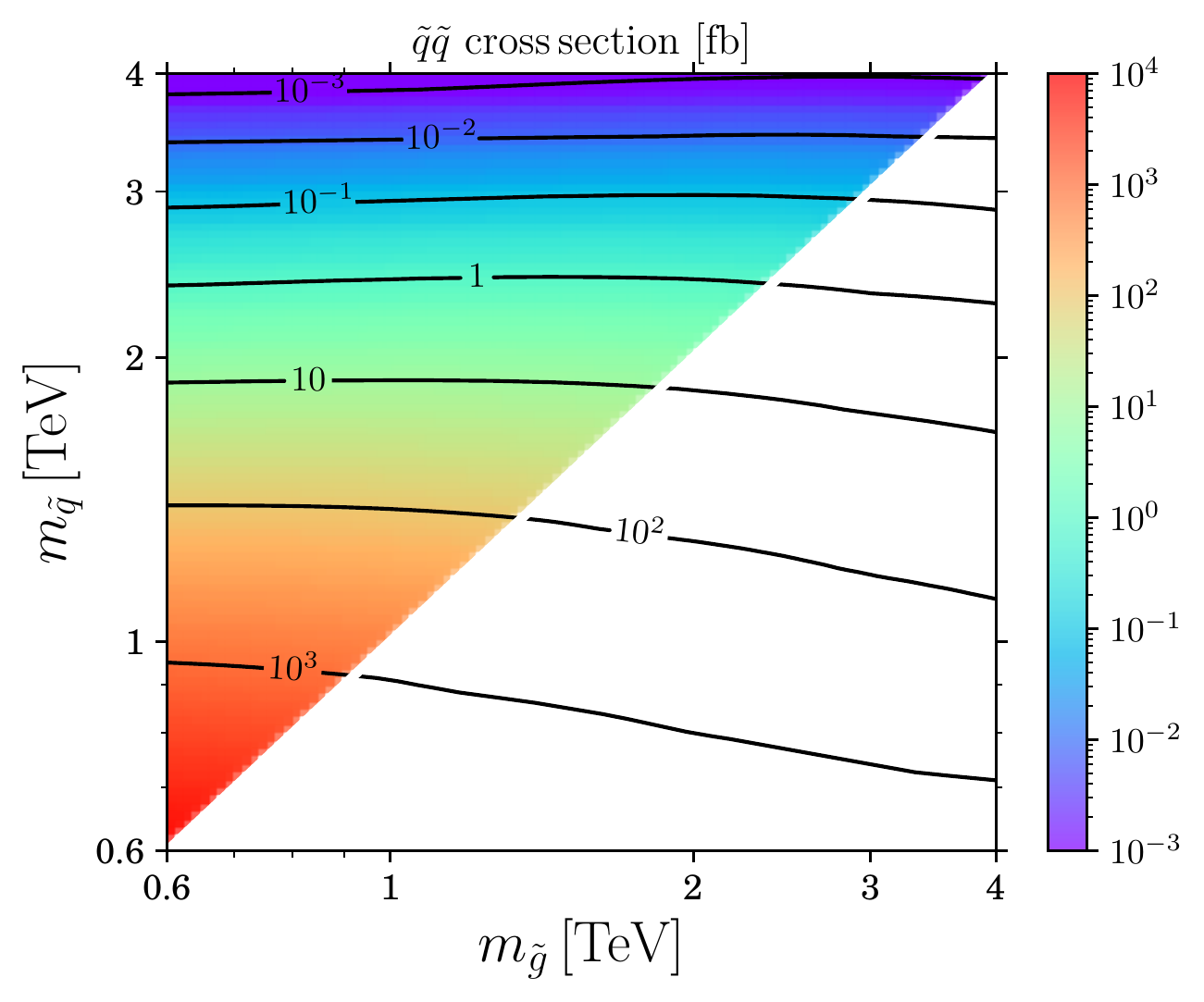}
      \includegraphics[width=0.48\textwidth]{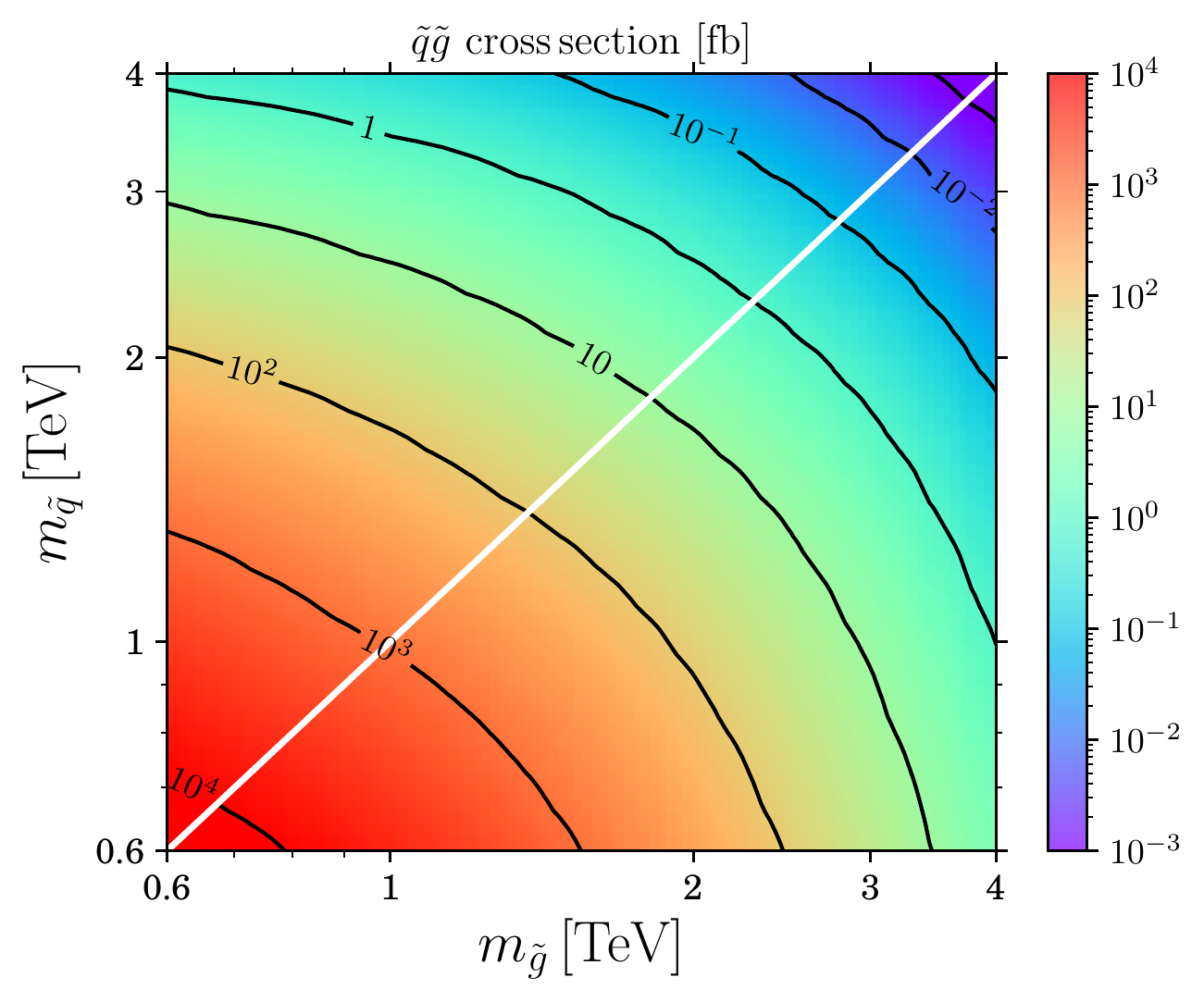}
\hspace{1mm}
      \includegraphics[width=0.48\textwidth]{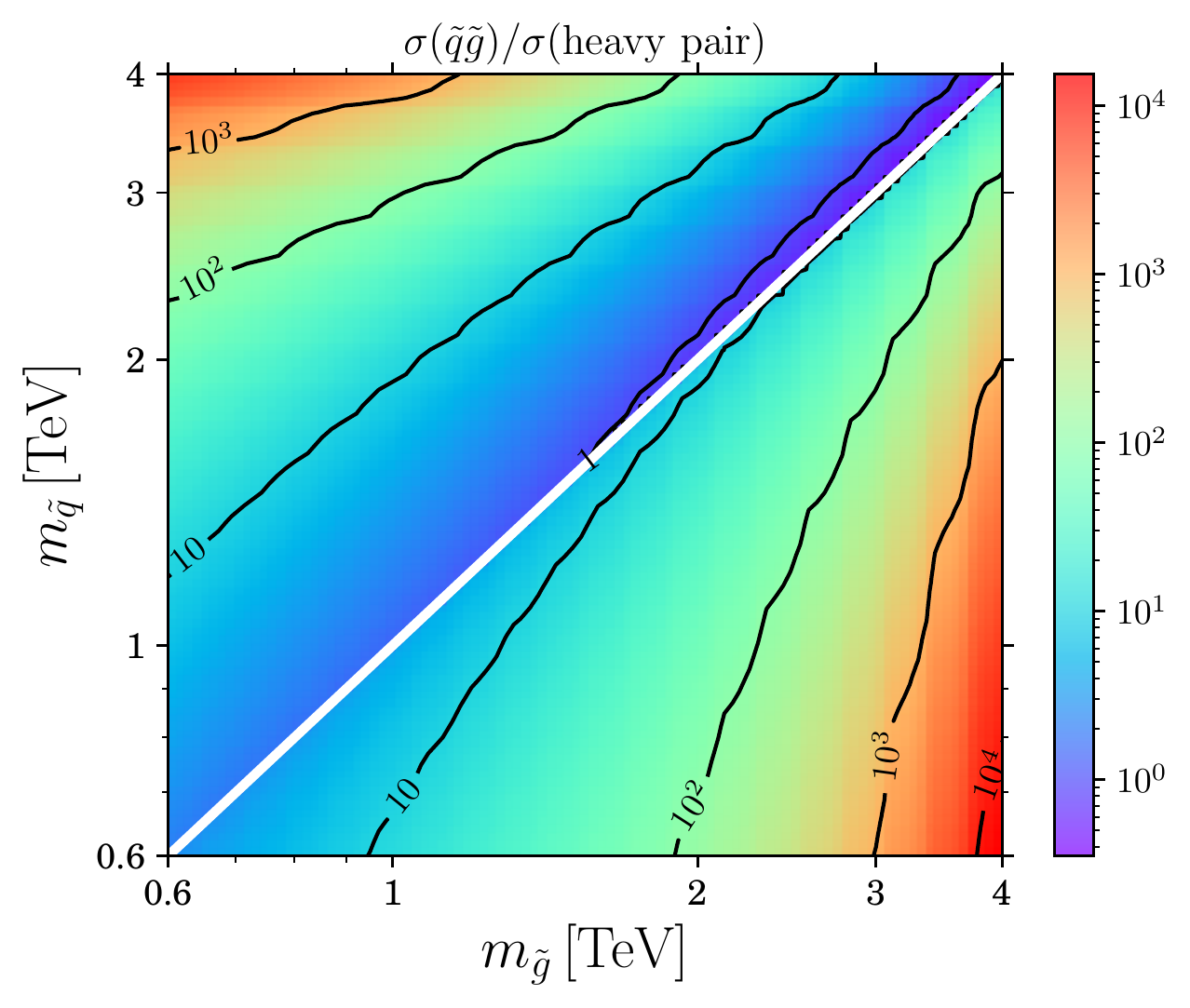}
\caption{\label{fig:QWxsec}
NLO cross sections of various production modes (and their ratios) of gluinos and squarks at the 13 TeV LHC are presented in the ($m_{\tilde g}, m_{\tilde q}$) plane in fb. 
In the top panels, the gluino-gluino (left) and squark-squark (right) pair production cross sections are shown.
In the top left plot, we highlight the $m_{\tilde g} > m_{\tilde q}$ region, for which the $pp \to \tilde g \tilde g$ production results in high $p_T$ jets from $\tilde g \to q \tilde q$, even if the lightest neutralino is mass degenerate with the lightest coloured particle.
Analogously, in the top right plot we highlight the $m_{\tilde q} > m_{\tilde g}$ region, for which the same final state originates the $pp \to \tilde q \tilde q$ process, followed by the $\tilde q \to q \tilde g$ decays.
The bottom left panel is for the associated squark-gluino production.
The bottom right panel shows the relative rate of the associated squark-gluino production cross section with respect to the gluino-gluino (in the lower right half with $m_{\tilde g} > m_{\tilde q}$)
and squark-squark (in the upper left half with $m_{\tilde q} > m_{\tilde g}$) pair productions cross section, respectively.
}
\label{fig:QGxsec}
\end{figure}

Figure~\ref{fig:QGxsec} shows NLO cross sections, calculated using {\tt Prospino}~\cite{Beenakker:1996ch}, of various production modes (and their ratios) of gluinos and squarks  at the 13 TeV LHC, presented in the ($m_{\tilde g}, m_{\tilde q}$) plane in fb. In the top panels, the gluino-gluino (left) and squark-squark (right) pair production cross sections are shown. In the top-left and top-right plots, the lower-right 
and upper-left regions correspond to 
({\bf a2}) and ({\bf b2}) processes, respectively,
and filled with colours.
In these regions, the produced coloured particle (gluino or squark) is heavier than the other and their decays,
$\tilde g \to q \tilde q$ or $\tilde q \to q \tilde g$,
produce high $p_T$ jets.
We see that the cross sections of the ({\bf a2}) and ({\bf b2}) processes diminishes rather quickly as the mass of the produced particle increases.
For example, the gluino-gluino (squark-squark) cross section decreases two orders of 
magnitude as the gluino (squark) mass increases by 1 TeV. 

The bottom left panel shows the NLO cross section of the associated squark-gluino production. When compared to the gluino (squark) pair production, it decreases much slower as a function of the gluino (squark) mass. Keeping the squark (gluino) mass fixed at $\sim 1$ TeV, the squark-gluino associated production cross section decreases an order of magnitude when the gluino (squark) mass is increased by 1 TeV. This is largely because the luminosity functions for the squark-gluino is larger than those for squark-squark and gluino-gluino in this mass range. We observe that as far as the lighter coloured particle is around 1 TeV, the production rate of $pp \to \tilde q \tilde g$ is sizeable ($\sim 10$ fb) even if the heavier coloured particle is around 3 TeV.

The bottom right panel shows the relative rate of the associated squark-gluino production with respect to the gluino-gluino (in the lower right half with $m_{\tilde g} > m_{\tilde q}$) and squark-squark (in the upper left half with $m_{\tilde q} > m_{\tilde g}$) pair productions, respectively. We see that the cross section of the associated production is almost always larger than that of the ({\bf a2}) and ({\bf b2}) processes. The relative rate of the associated production enhances particularly in the hierarchical mass regions ($m_{\tilde g} \gg m_{\tilde q}$ and $m_{\tilde q} \gg m_{\tilde g}$).
For example, around $(m_{\tilde g}, m_{\tilde q}) \simeq (1, 3)$ TeV, $\sigma(\tilde q \tilde g)/\sigma(\tilde q \tilde q) \sim 500$. Similarly, $\sigma(\tilde q \tilde g)/\sigma(\tilde g \tilde g) \sim 700$ around $(m_{\tilde g}, m_{\tilde q}) \simeq (3, 1)$ TeV.

\begin{figure}[t!]
\centering
      \includegraphics[width=0.49\textwidth]{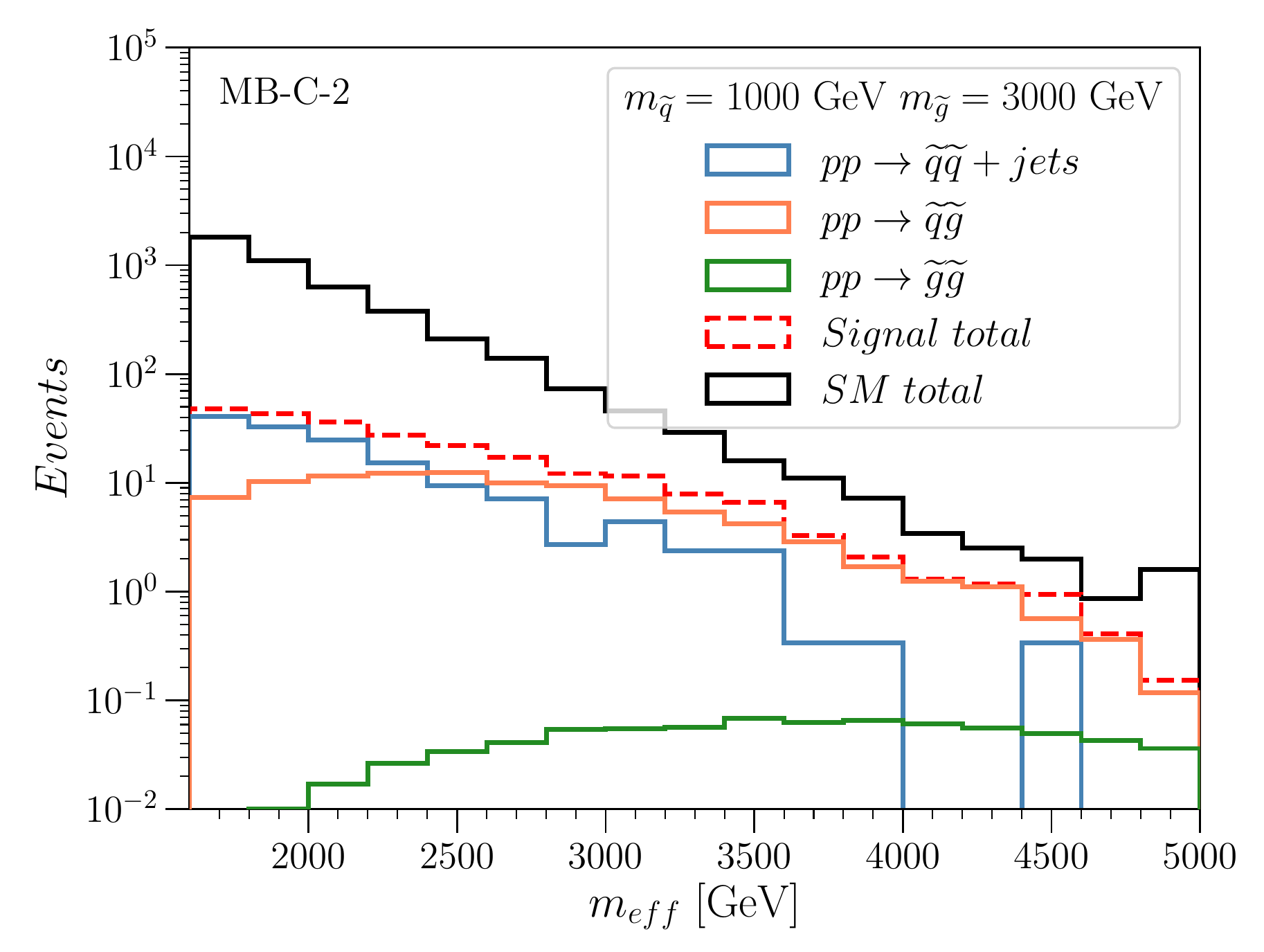}
      \includegraphics[width=0.49\textwidth]{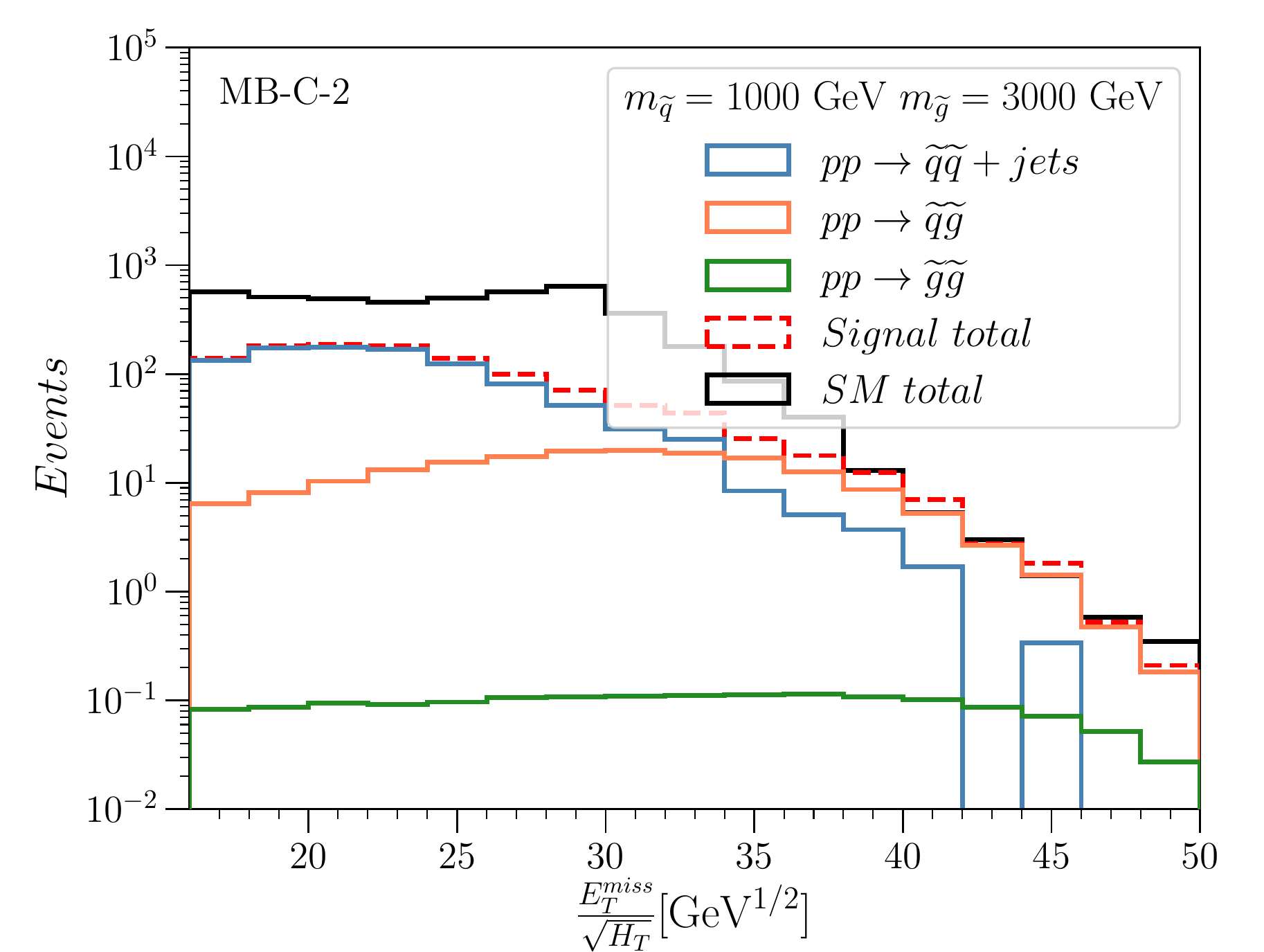}
      \includegraphics[width=0.49\textwidth]{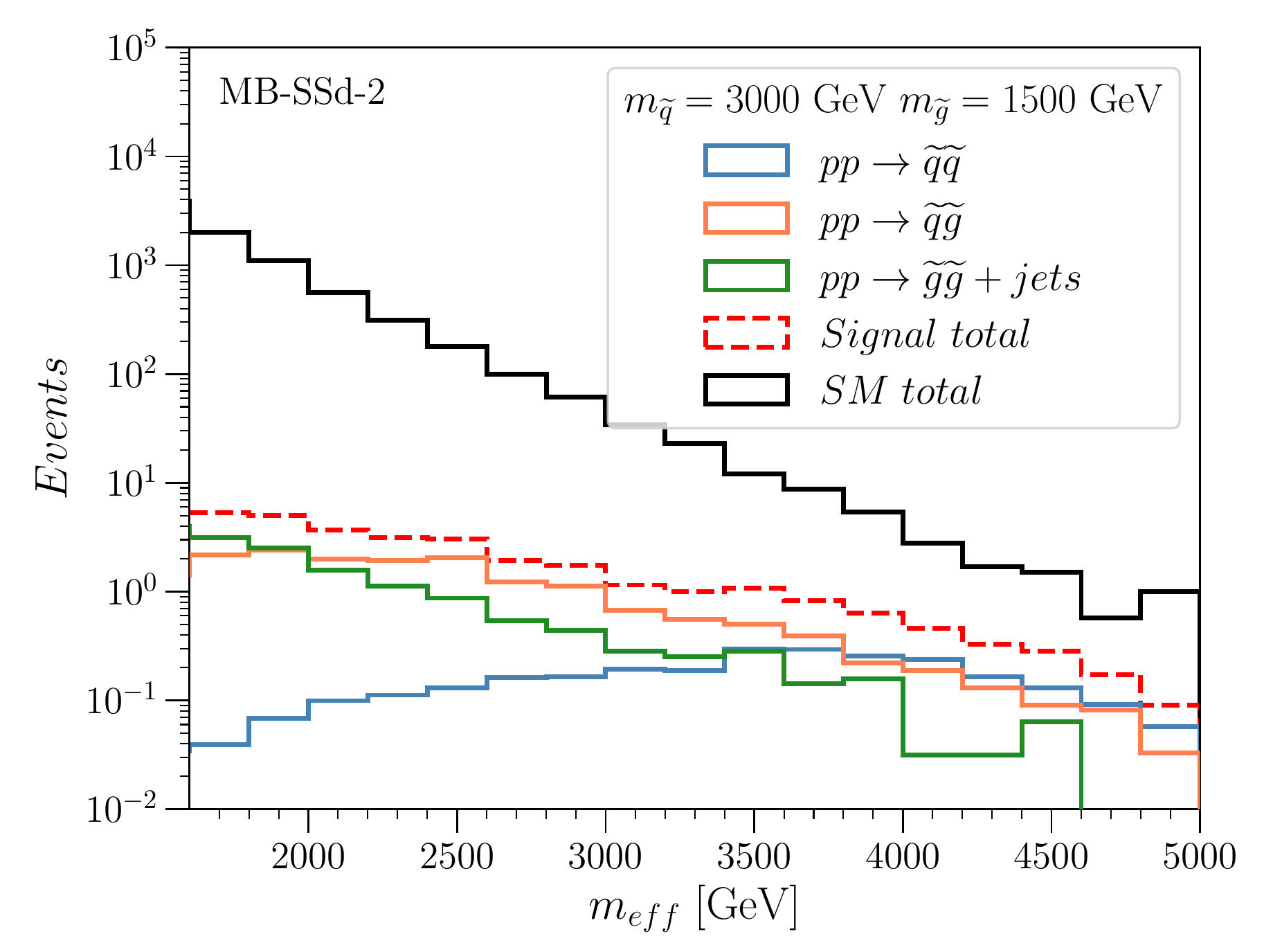}
      \includegraphics[width=0.49\textwidth]{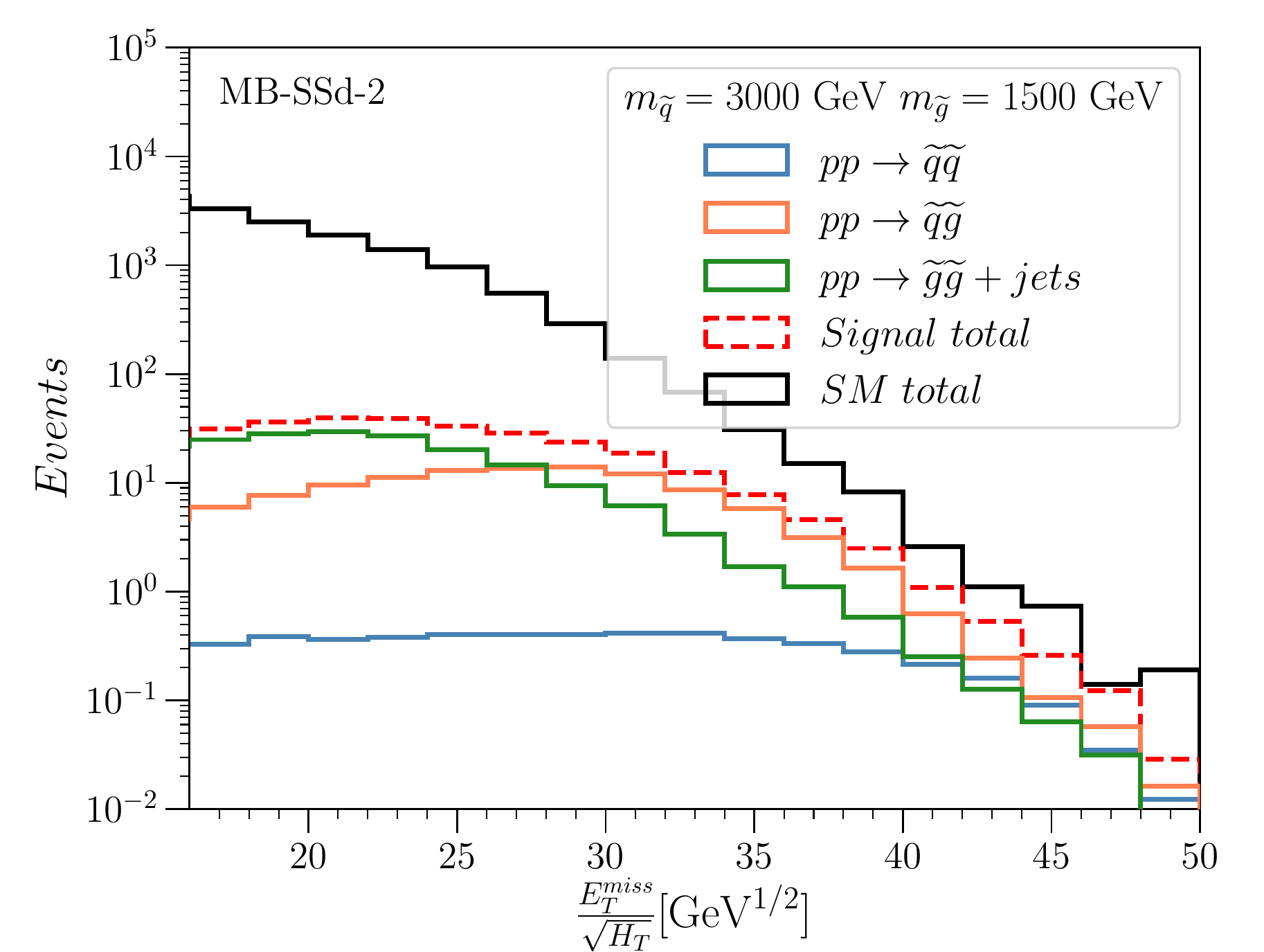}
\caption{\label{fig:Meff_QG}
The $m_{\rm eff}$ (left) and $E_T^{\rm miss}/\sqrt{H_T}$ (right) distributions after imposing the remaining cuts in the most sensitive signal region 
of the ATLAS ``jets + $\met$'' analysis \cite{ATLAS:2020syg}
for the given mass point.
In the upper plots, the masses are taken to be $(m_{\tilde q}, m_{\tilde g}) = (1.0, 3.0)\,{\rm TeV}$,
whereas they are
$(m_{\tilde q}, m_{\tilde g}) = (3.0, 1.5)\,{\rm TeV}$
in the lower plots.
The SM background distributions are taken from the ATLAS ``jets + $\met$'' paper \cite{ATLAS:2020syg}.
}
\end{figure}

The processes ({\bf a3}) and ({\bf b3}) are pair productions of lighter coloured particles associated with the initial state QCD radiation.
In order for these processes to contribute to the signal regions,
$\jQCDs$ have to be energetic 
and result in a large $E_T^{\rm miss}$.
To see the impact of these processes in the event selection, 
we compare in Fig.~\ref{fig:Meff_QG} 
the contributions from
the three signal sub-processes
to the distributions of the $m_{\rm eff}$ (left) and $E_T^{\rm miss}/\sqrt{H_T}$ (right) variables.
In the upper panels,
the masses are taken to be
$(m_{\tilde q}, m_{\tilde g}) = (1, 3)\,{\rm TeV}$
and
the three signal sub-processes,
({\bf a1}), ({\bf a2}) and ({\bf a3}),
are shown
together with the sum of these three (dashed-red)
and the total SM contribution (solid-black), which is
taken from the ATLAS ``jets + $\met$'' paper \cite{ATLAS:2020syg}.
The distributions are made 
after imposing the MB-C-2 event selection,
which is the most sensitive signal region 
for the chosen mass point.
We see that the main signal contributions come from
the associated squark-gluino production, $pp \to \tilde q \tilde g$,
and the pair production of lighter coloured particles, $pp \to \tilde q \tilde q + \jQCDs$,
for both $m_{\rm eff}$ and 
$E_T^{\rm miss}/\sqrt{H_T}$ distributions.
In particular, at high values of 
$m_{\rm eff}$ and $E_T^{\rm miss}/\sqrt{H_T}$,
the associated production 
dominates over the pair production.
The same feature is observed 
in the bottom 
two plots in Fig.\ \ref{fig:Meff_QG}
where the masses are 
taken to be
$(m_{\tilde q}, m_{\tilde g}) = (3.0, 1.5)\,{\rm TeV}$.
For this mass point the most sensitive signal region is MB-SSd-2.
At low values of $m_{\rm eff}$
and $E_T^{\rm miss}/\sqrt{H_T}$,
the associated production, $pp \to \tilde q \tilde g$,
and the pair production of ligher coloured particles, $pp \to \tilde g \tilde g + \jQCDs$,
dominate,
whereas at high values  
the associated production as well as the pair production of heavier coloured particles, $pp \to \tilde q \tilde q$, give the main contributions.

\begin{figure}[t!]
\centering
      \includegraphics[width=0.8\textwidth]{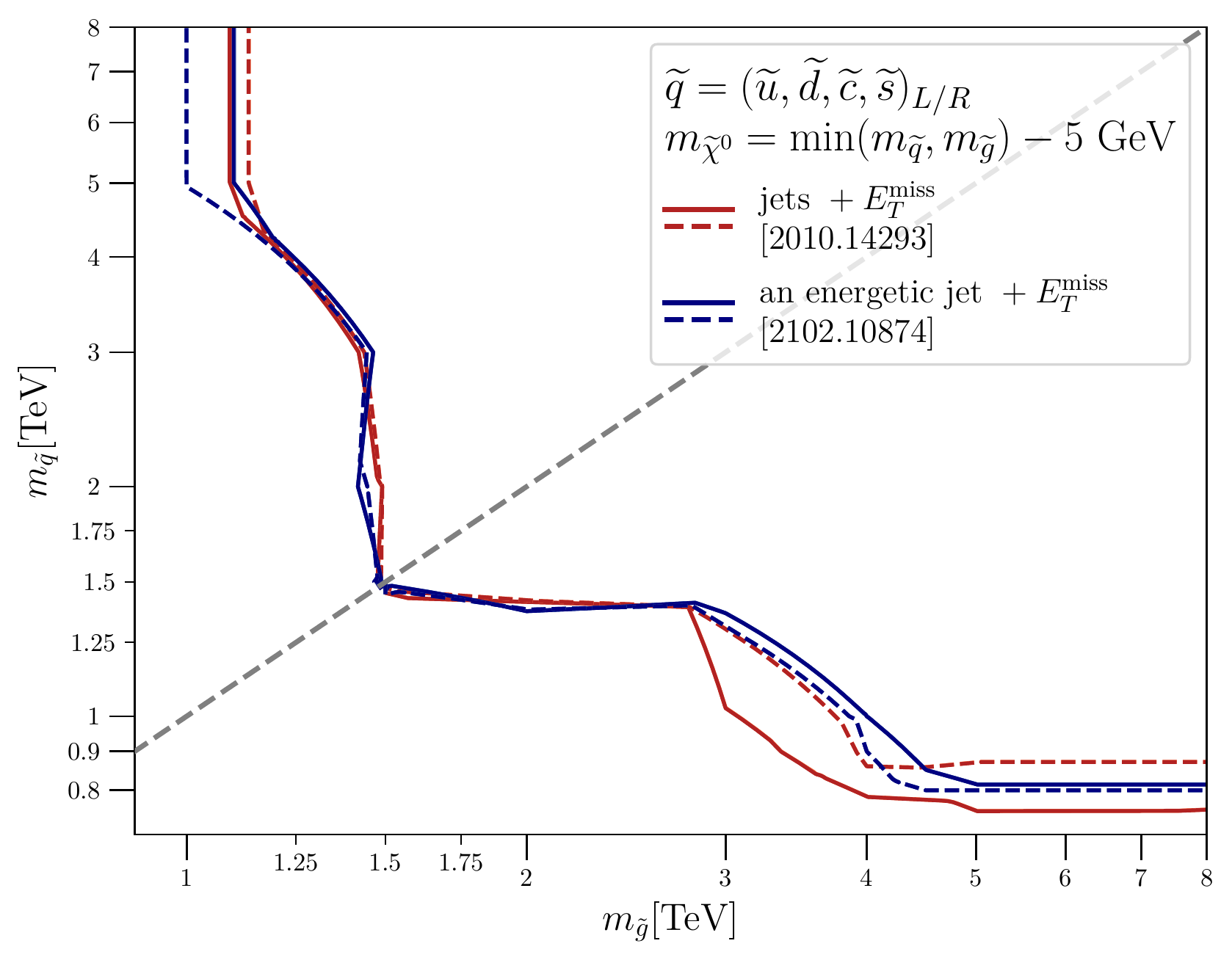}
\caption{The exclusion limit in the ($m_{\tilde g}$, $m_{\tilde q}$) plane with constraint $m_{\tilde \chi_1^0} = {\rm min}(m_{\tilde g}, m_{\tilde q})-5$ GeV. The blue and red contours represent the limits obtained from the ATLAS ``an energetic jet + $\met$'' \cite{ATLAS:2021kxv} and ``jets + $\met$'' \cite{ATLAS:2020syg} analyses, respectively. The solid and dashed contours correspond to the observed and expected limits, respectively. The diagonal grey dashed line indicates $m_{\tilde g} = m_{\tilde q}$. }
\label{fig:GQ_limit}
\end{figure}

Figure \ref{fig:GQ_limit} displays the exclusion limit in the ($m_{\tilde g}$, $m_{\tilde q}$) plane obtained by recasting ``an energetic jet + $\met$'' \cite{ATLAS:2021kxv} (blue) and ``jets + $\met$'' (red) \cite{ATLAS:2020syg} analyses. The solid and dashed contours correspond to the observed and expected exclusion limits, respectively. For each mass point the lightest neutralino mass is fixed at  $m_{\tilde \chi_1^0} = {\rm min}(m_{\tilde g}, m_{\tilde q}) - 5\,{\rm GeV}$ so that  the lighter coloured particle is approximately mass degenerate with $\tilde \chi_1^0$. Among many signal regions defined in the ``jets + $\met$'' analysis \cite{ATLAS:2020syg} only the signal regions MB-SSd-2 and MB-C (see Table \ref{tb:cut}) contribute to the final exclusion contours. This is because signal events typically have a small number of high $p_T$ jets since decays of the lighter coloured particle do not produce energetic particles due to the mass requirement,  $m_{\tilde \chi_1^0} = {\rm min}(m_{\tilde g}, m_{\tilde q}) - 5\,{\rm GeV}$.

We observe that the ``an energetic jet + $\met$''  and ``jets + $\met$'' analyses provide very similar exclusion contours on the plane. The only exception is the region around $(m_{\tilde g}, m_{\tilde q}) \sim (3.5, 1.0)$ TeV, where the observed limit from ``jets + $\met$'' is slightly weaker than that from ``an energetic jet + $\met$''.  We checked that this is due to a mild excess  observed in the MB-C-2 signal region.

In the hierarchical mass regions, $m_{\tilde g} \gg m_{\tilde q}$ and $m_{\tilde q} \gg m_{\tilde g}$, only $pp \to \tilde q \tilde q + \jQCDs$ and $pp \to \tilde g \tilde g + \jQCDs$ processes contribute, respectively. In these regions, squarks are excluded up to $\sim 0.8$ TeV, independently of the gluino mass, whereas gluino lighter than $\sim 1.2$ TeV is excluded regardless of the squark mass.
In a moderately hierarchical mass region, $m_{\tilde g} \gtrsim m_{\tilde q}$ ($m_{\tilde q} \gtrsim m_{\tilde g}$), all three signal sub-processes, ({\bf a1}), ({\bf a2}) and ({\bf a3}) (({\bf b1}), ({\bf b2}) and ({\bf b3})), contribute. At $m_{\tilde q} = 3$ TeV, the gluino is excluded up to 1.5 TeV. By lowering squark mass  below 3 TeV, the production rates of $pp \to \tilde q \tilde q$ and $pp \to \tilde q \tilde g$ increase,  while the acceptance decreases since the mass difference $m_{\tilde q} - m_{\tilde g}$ gets smaller and jets from the $\tilde q \to q \tilde g$ become less energetic. Due to cancellation of these two effects, the gluino mass limit, $\sim 1.5$ TeV, stays constant between $m_{\tilde q} \sim [1.4,\, 3.0]$ TeV range. The same feature is observed for the $m_{\tilde g} \gtrsim m_{\tilde q}$ region. Squarks are excluded up to $\sim 1.4$ TeV, regardless of the gluino mass in the $m_{\tilde g} \sim [1.5,\,3.0]$ TeV range.
In the $m_{\tilde g} \simeq m_{\tilde q}$ region,
decays of coloured particles cannot produce energetic jets due to a compressed mass spectrum. The dominant contribution to the signal regions comes from the $pp \to \tilde q \tilde q + \jQCDs$ and $pp \to \tilde g \tilde g + \jQCDs$ processes. The lower mass limit $\sim 1.5$ TeV is imposed on ${\rm min}(m_{\tilde q}, m_{\tilde g})$ in this region.

%%%%%%%%%%%%%%%%%%%%%%%%%%%%%%%%%%%%%%%%%
\section{Conclusions}
\label{sec:concl}
%%%%%%%%%%%%%%%%%%%%%%%%%%%%%%%%%%%%%%%%%

Mono- and di-jet channels are a powerful tool to look for the production of (effectively) invisible and stable particles at the LHC. In this paper, we have studied two distinct SUSY scenarios, where a single high $p_T$ jet originates from SUSY particle pair production, to which mono- and di-jet event selections are particularly sensitive.

The first scenario is effectively described by a squark-wino simplified model, where the mass hierarchy, (a few TeV)$\, > m_{\tilde q} \gg m_{\widetilde W}$, is assumed. 
We found that in this scenario one cannot neglect the contribution of the associated squark-wino production when deriving the limits.
Ignoring the associated production can result in severely underestimated bounds on the model. 
The cross section of the associated production is larger than of squark pair production already for $m_{\tilde q} \gtrsim 500 - 800$ GeV, depending on the effective number of squarks accessible by the collider energy.
Furthermore $pp \to \widetilde W \widetilde W + \jQCDs$ also contributes to the relevant signal regions. 
Comparing the exclusion limit derived only from
the $pp \to \tilde q \tilde q^*$ mode
and that from all three signal sub-processes including
$pp \to \tilde q \widetilde W$ and $\wino \wino + \jQCDs$,
we found that
the exclusion power is significantly enhanced for the latter case. The largest effect is observed for the case when just the first generation left-handed squarks are kinematically accessible. For example, the lower limit on the squark mass at $m_{\wino} = 400$ GeV is extended by $\sim 45$, 20 and 10 \% in the 2-, 4- and 8-flavour squark scenarios, respectively. 
Finally, we note that the similar effect will be present for the squark-bino model, although the threshold, above which the contribution from the associated production becomes prominent, is shifted to $m_{\tilde q} \sim 2$ TeV. 
With the current integrated luminosity this subprocess can be neglected for the squark-bino model. 
However, it should be reconsidered for the analyses at the HL-LHC with an integrated luminosity 3000~fb$^{-1}$.

The second analysed scenario is the gluino-squark simplified model with the bino-like LSP neutralino, where the neutralino is almost mass degenerate with the lighter of the gluino or squark, 
as required in the gluino-bino and squark-bino coannihilation scenarios.

In this scenario, high $p_T$ jets can only be produced from decays of the heavier coloured SUSY particle or the initial state QCD radiation. We studied the cross sections and kinematical distributions of the three main signal sub-processes.
By recasting the ATLAS ``an energetic jet + $\met$'' \cite{ATLAS:2021kxv} and ``jets + $\met$'' \cite{ATLAS:2020syg} analyses, the current exclusion limit  has been derived on the ($m_{\tilde g}, m_{\tilde q}$) plane,  fixing the neutralino mass at $m_{\tilde \chi_1^0} = {\rm min}(m_{\tilde q}, m_{\tilde g}) - 5$ GeV for each mass point. We have seen that the two ATLAS analyses have a similar performance and excluded the squarks (gluino) up to $\sim 1$ (1.2) TeV for very heavy gluino (squarks). If the gluino-squark masses are of the similar order, the lower limit on the squark (gluino) mass is given by $\sim 1.5$ TeV for the $m_{\tilde g / \tilde q} \sim (1.5 - 3)$ TeV range (see Fig.\ref{fig:GQ_limit}).

%{\color{red} Although we used the ATLAS analyses to demonstrate the effect of associated productions on the mass limit, we expect the similar effect is present in the corresponding CMS analyses.}

\section*{Acknowledgments}

The research leading to these results has received funding from the Norwegian Financial Mechanism 2014-2021, grant 2019/34/H/ST2/00707. The work of I.L.\ was directly funded from the grant.
The work of K.S.\ is partially supported by the National Science Centre, Poland, under research grant 2017/26/E/ST2/00135. The work of R.M.\ is partially supported by the National Science Centre, Poland, under research grant Preludium20 2021/41/N/ST2/00972. MN is supported by Grant-in-Aid for Scientific Research on Innovative Areas(16H06492) JSPS KAK-ENHI 22K03629. K.R.\ was partially supported by the National Science Centre, Poland under grants: 2018/31/B/ST2/02283, 2019/35/B/ST2/02008.

% \appendix
% \section{}
% \label{ap:}

\bibliography{refs}

\providecommand{\href}[2]{#2}\begingroup\begin{thebibliography}{10}

\bibitem{Farrar:1978xj}
G.~R.~Farrar and P.~Fayet, ``{Phenomenology of the Production, Decay, and
  Detection of New Hadronic States Associated with Supersymmetry},''
  \href{https://doi.org/10.1016/0370-2693(78)90858-4}{Phys.\  Lett.\  B
  {\bfseries 76} (1978) 575--579}.

\bibitem{Weinberg:1982tp}
S.~Weinberg, ``{Upper Bound on Gauge Fermion Masses},''
  \href{https://doi.org/10.1103/PhysRevLett.50.387}{Phys.\  Rev.\  Lett.\
  {\bfseries 50} (1983) 387}.

\bibitem{Ellis:1983ew}
M.~A.~Srednicki, ed., ``{Supersymmetric Relics from the Big Bang},''
  \href{https://doi.org/10.1016/0550-3213(84)90461-9}{Nucl.\  Phys.\  B
  {\bfseries 238} (1984) 453--476}.

\bibitem{Alwall:2008va}
J.~Alwall, M.-P.~Le, M.~Lisanti, and J.~G.~Wacker, ``{Model-Independent Jets
  plus Missing Energy Searches},''
  \href{https://doi.org/10.1103/PhysRevD.79.015005}{Phys.\  Rev.\  D {\bfseries
  79} (2009) 015005} {\ttfamily
  [\href{https://arxiv.org/abs/0809.3264}{arXiv:0809.3264}]}.

\bibitem{ATLAS:2021kxv}
{\bfseries ATLAS} Collaboration, ``{Search for new phenomena in events with an
  energetic jet and missing transverse momentum in $pp$ collisions at $\sqrt
  {s}$ =13 TeV with the ATLAS detector},''
  \href{https://doi.org/10.1103/PhysRevD.103.112006}{Phys.\  Rev.\  D
  {\bfseries 103} (2021) 112006} {\ttfamily
  [\href{https://arxiv.org/abs/2102.10874}{arXiv:2102.10874}]}.

\bibitem{CMS:2021far}
{\bfseries CMS} Collaboration, ``{Search for new particles in events with
  energetic jets and large missing transverse momentum in proton-proton
  collisions at $ \sqrt{s} $ = 13 TeV},''
  \href{https://doi.org/10.1007/JHEP11(2021)153}{JHEP {\bfseries 11} (2021)
  153} {\ttfamily [\href{https://arxiv.org/abs/2107.13021}{arXiv:2107.13021}]}.

\bibitem{ATLAS-CONF-2011-096}
{\bfseries ATLAS} Collaboration, ``{Search for New Phenomena in Monojet plus
  Missing Transverse Momentum Final States using 1~fb$^{-1}$ of $pp$ Collisions
  at $\sqrt{s}=7$ TeV with the ATLAS Detector},''
  \href{https://cds.cern.ch/record/1369187}{ATLAS--CONF--2011--096}, CERN,
  Geneva, 2011.

\bibitem{CMS-PAS-EXO-11-059}
{\bfseries CMS} Collaboration, ``{Search for New Physics with a Monojet and
  Missing Transverse Energy in $pp$ Collisions at $\sqrt{s} = 7$ TeV},''
  \href{https://cds.cern.ch/record/1376675}{CMS--PAS--EXO--11--059}, CERN,
  Geneva, 2011.

\bibitem{Dreiner:2012gx}
H.~K.~Dreiner, M.~Kramer, and J.~Tattersall, ``{How low can SUSY go? Matching,
  monojets and compressed spectra},''
  \href{https://doi.org/10.1209/0295-5075/99/61001}{EPL {\bfseries 99} (2012)
  61001} {\ttfamily [\href{https://arxiv.org/abs/1207.1613}{arXiv:1207.1613}]}.

\bibitem{ATLAS:2020syg}
{\bfseries ATLAS} Collaboration, ``{Search for squarks and gluinos in final
  states with jets and missing transverse momentum using 139 fb$^{-1}$ of
  $\sqrt{s}$ =13 TeV $pp$ collision data with the ATLAS detector},''
  \href{https://doi.org/10.1007/JHEP02(2021)143}{JHEP {\bfseries 02} (2021)
  143} {\ttfamily [\href{https://arxiv.org/abs/2010.14293}{arXiv:2010.14293}]}.

\bibitem{DeSimone:2014qkh}
A.~De~Simone, G.~F.~Giudice, and A.~Strumia, ``{Benchmarks for Dark Matter
  Searches at the LHC},'' \href{https://doi.org/10.1007/JHEP06(2014)081}{JHEP
  {\bfseries 06} (2014) 081} {\ttfamily
  [\href{https://arxiv.org/abs/1402.6287}{arXiv:1402.6287}]}.

\bibitem{Profumo:2004wk}
S.~Profumo and C.~E.~Yaguna, ``{Gluino coannihilations and heavy bino dark
  matter},'' \href{https://doi.org/10.1103/PhysRevD.69.115009}{Phys.\  Rev.\  D
  {\bfseries 69} (2004) 115009} {\ttfamily
  [\href{https://arxiv.org/abs/hep-ph/0402208}{hep-ph/0402208}]}.

\bibitem{Harigaya:2014dwa}
K.~Harigaya, K.~Kaneta, and S.~Matsumoto, ``{Gaugino coannihilations},''
  \href{https://doi.org/10.1103/PhysRevD.89.115021}{Phys.\  Rev.\  D {\bfseries
  89} (2014) 115021} {\ttfamily
  [\href{https://arxiv.org/abs/1403.0715}{arXiv:1403.0715}]}.

\bibitem{Ellis:2015vaa}
J.~Ellis, F.~Luo, and K.~A.~Olive, ``{Gluino Coannihilation Revisited},''
  \href{https://doi.org/10.1007/JHEP09(2015)127}{JHEP {\bfseries 09} (2015)
  127} {\ttfamily [\href{https://arxiv.org/abs/1503.07142}{arXiv:1503.07142}]}.

\bibitem{CMS:2019zmd}
{\bfseries CMS} Collaboration, ``{Search for supersymmetry in proton-proton
  collisions at 13 TeV in final states with jets and missing transverse
  momentum},'' \href{https://doi.org/10.1007/JHEP10(2019)244}{JHEP {\bfseries
  10} (2019) 244} {\ttfamily
  [\href{https://arxiv.org/abs/1908.04722}{arXiv:1908.04722}]}.

\bibitem{ATLAS:2022rme}
{\bfseries ATLAS} Collaboration, ``{Search for long-lived charginos based on a
  disappearing-track signature using 136 fb$^{-1}$ of pp collisions at
  $\sqrt{s}$~=~13~TeV with the ATLAS detector},''
  \href{https://doi.org/10.1140/epjc/s10052-022-10489-5}{Eur.\  Phys.\  J.\  C
  {\bfseries 82} (2022) 606} {\ttfamily
  [\href{https://arxiv.org/abs/2201.02472}{arXiv:2201.02472}]}.

\bibitem{Buanes:2022wgm}
T.~Buanes, I.~n.~Lara, K.~Rolbiecki, and K.~Sakurai, ``{LHC constraints on
  electroweakino dark matter revisited}.'' {\ttfamily
  \href{https://arxiv.org/abs/2208.04342}{arXiv:2208.04342}}.

\bibitem{Fuks:2013vua}
B.~Fuks, M.~Klasen, D.~R.~Lamprea, and M.~Rothering, ``{Precision predictions
  for electroweak superpartner production at hadron colliders with
  Resummino},'' \href{https://doi.org/10.1140/epjc/s10052-013-2480-0}{Eur.\
  Phys.\  J.\  C {\bfseries 73} (2013) 2480} {\ttfamily
  [\href{https://arxiv.org/abs/1304.0790}{arXiv:1304.0790}]}.

\bibitem{Fiaschi:2022odp}
J.~Fiaschi, B.~Fuks, M.~Klasen, and A.~Neuwirth, ``{Soft gluon resummation for
  associated squark-electroweakino production at the LHC},''
  \href{https://doi.org/10.1007/JHEP06(2022)130}{JHEP {\bfseries 06} (2022)
  130} {\ttfamily [\href{https://arxiv.org/abs/2202.13416}{arXiv:2202.13416}]}.

\bibitem{Beenakker:1996ch}
W.~Beenakker, R.~Hopker, M.~Spira, and P.~M.~Zerwas, ``{Squark and gluino
  production at hadron colliders},''
  \href{https://doi.org/10.1016/S0550-3213(97)80027-2}{Nucl.\  Phys.\  B
  {\bfseries 492} (1997) 51--103} {\ttfamily
  [\href{https://arxiv.org/abs/hep-ph/9610490}{hep-ph/9610490}]}.

\bibitem{Kulesza:2008jb}
A.~Kulesza and L.~Motyka, ``{Threshold resummation for squark-antisquark and
  gluino-pair production at the LHC},''
  \href{https://doi.org/10.1103/PhysRevLett.102.111802}{Phys.\  Rev.\  Lett.\
  {\bfseries 102} (2009) 111802} {\ttfamily
  [\href{https://arxiv.org/abs/0807.2405}{arXiv:0807.2405}]}.

\bibitem{Kulesza:2009kq}
A.~Kulesza and L.~Motyka, ``{Soft gluon resummation for the production of
  gluino-gluino and squark-antisquark pairs at the LHC},''
  \href{https://doi.org/10.1103/PhysRevD.80.095004}{Phys.\  Rev.\  D {\bfseries
  80} (2009) 095004} {\ttfamily
  [\href{https://arxiv.org/abs/0905.4749}{arXiv:0905.4749}]}.

\bibitem{Beenakker:2009ha}
W.~Beenakker, S.~Brensing, M.~Kramer, A.~Kulesza, \emph{et al}., ``{Soft-gluon
  resummation for squark and gluino hadroproduction},''
  \href{https://doi.org/10.1088/1126-6708/2009/12/041}{JHEP {\bfseries 12}
  (2009) 041} {\ttfamily
  [\href{https://arxiv.org/abs/0909.4418}{arXiv:0909.4418}]}.

\bibitem{Beenakker:2011sf}
W.~Beenakker, S.~Brensing, M.~Kramer, A.~Kulesza, \emph{et al}., ``{NNLL
  resummation for squark-antisquark pair production at the LHC},''
  \href{https://doi.org/10.1007/JHEP01(2012)076}{JHEP {\bfseries 01} (2012)
  076} {\ttfamily [\href{https://arxiv.org/abs/1110.2446}{arXiv:1110.2446}]}.

\bibitem{Beenakker:2013mva}
W.~Beenakker, T.~Janssen, S.~Lepoeter, M.~Kr\"amer, \emph{et al}., ``{Towards
  NNLL resummation: hard matching coefficients for squark and gluino
  hadroproduction},'' \href{https://doi.org/10.1007/JHEP10(2013)120}{JHEP
  {\bfseries 10} (2013) 120} {\ttfamily
  [\href{https://arxiv.org/abs/1304.6354}{arXiv:1304.6354}]}.

\bibitem{Beenakker:2014sma}
W.~Beenakker, C.~Borschensky, M.~Kr\"amer, A.~Kulesza, \emph{et al}., ``{NNLL
  resummation for squark and gluino production at the LHC},''
  \href{https://doi.org/10.1007/JHEP12(2014)023}{JHEP {\bfseries 12} (2014)
  023} {\ttfamily [\href{https://arxiv.org/abs/1404.3134}{arXiv:1404.3134}]}.

\bibitem{Beenakker:2016lwe}
W.~Beenakker, C.~Borschensky, M.~Kr\"amer, A.~Kulesza, and E.~Laenen,
  ``{NNLL-fast: predictions for coloured supersymmetric particle production at
  the LHC with threshold and Coulomb resummation},''
  \href{https://doi.org/10.1007/JHEP12(2016)133}{JHEP {\bfseries 12} (2016)
  133} {\ttfamily [\href{https://arxiv.org/abs/1607.07741}{arXiv:1607.07741}]}.

\bibitem{madgraph5}
J.~Alwall, R.~Frederix, S.~Frixione, V.~Hirschi, \emph{et al}., ``{The
  automated computation of tree-level and next-to-leading order differential
  cross sections, and their matching to parton shower simulations},''
  \href{https://doi.org/10.1007/JHEP07(2014)079}{JHEP {\bfseries 07} (2014)
  079} {\ttfamily [\href{https://arxiv.org/abs/1405.0301}{arXiv:1405.0301}]}.

\bibitem{Ball:2012cx}
R.~D.~Ball \emph{et al}., ``{Parton distributions with LHC data},''
  \href{https://doi.org/10.1016/j.nuclphysb.2012.10.003}{Nucl.\  Phys.\  B
  {\bfseries 867} (2013) 244--289} {\ttfamily
  [\href{https://arxiv.org/abs/1207.1303}{arXiv:1207.1303}]}.

\bibitem{Buckley:2014ana}
A.~Buckley, J.~Ferrando, S.~Lloyd, K.~Nordstr\"om, \emph{et al}., ``{LHAPDF6:
  parton density access in the LHC precision era},''
  \href{https://doi.org/10.1140/epjc/s10052-015-3318-8}{Eur.\  Phys.\  J.\  C
  {\bfseries 75} (2015) 132} {\ttfamily
  [\href{https://arxiv.org/abs/1412.7420}{arXiv:1412.7420}]}.

\bibitem{Sjostrand:2014zea}
T.~Sj\"ostrand, S.~Ask, J.~R.~Christiansen, R.~Corke, \emph{et al}., ``{An
  introduction to PYTHIA 8.2},''
  \href{https://doi.org/10.1016/j.cpc.2015.01.024}{Comput.\  Phys.\  Commun.\
  {\bfseries 191} (2015) 159--177} {\ttfamily
  [\href{https://arxiv.org/abs/1410.3012}{arXiv:1410.3012}]}.

\bibitem{Mangano:2006rw}
M.~L.~Mangano, M.~Moretti, F.~Piccinini, and M.~Treccani, ``{Matching matrix
  elements and shower evolution for top-quark production in hadronic
  collisions},'' \href{https://doi.org/10.1088/1126-6708/2007/01/013}{JHEP
  {\bfseries 01} (2007) 013} {\ttfamily
  [\href{https://arxiv.org/abs/hep-ph/0611129}{hep-ph/0611129}]}.

\bibitem{Cacciari:2011ma}
M.~Cacciari, G.~P.~Salam, and G.~Soyez, ``{FastJet User Manual},''
  \href{https://doi.org/10.1140/epjc/s10052-012-1896-2}{Eur.\  Phys.\  J.\  C
  {\bfseries 72} (2012) 1896} {\ttfamily
  [\href{https://arxiv.org/abs/1111.6097}{arXiv:1111.6097}]}.

\bibitem{Cacciari:2005hq}
M.~Cacciari and G.~P.~Salam, ``{Dispelling the $N^{3}$ myth for the $k_t$
  jet-finder},'' \href{https://doi.org/10.1016/j.physletb.2006.08.037}{Phys.\
  Lett.\  B {\bfseries 641} (2006) 57--61} {\ttfamily
  [\href{https://arxiv.org/abs/hep-ph/0512210}{hep-ph/0512210}]}.

\bibitem{deFavereau:2013fsa}
{\bfseries DELPHES 3} Collaboration, ``{DELPHES 3, A modular framework for fast
  simulation of a generic collider experiment},''
  \href{https://doi.org/10.1007/JHEP02(2014)057}{JHEP {\bfseries 02} (2014)
  057} {\ttfamily [\href{https://arxiv.org/abs/1307.6346}{arXiv:1307.6346}]}.

\bibitem{checkmate1}
M.~Drees, H.~Dreiner, D.~Schmeier, J.~Tattersall, and J.~S.~Kim, ``{CheckMATE:
  Confronting your Favourite New Physics Model with LHC Data},''
  \href{https://doi.org/10.1016/j.cpc.2014.10.018}{Comput.\  Phys.\  Commun.\
  {\bfseries 187} (2015) 227--265} {\ttfamily
  [\href{https://arxiv.org/abs/1312.2591}{arXiv:1312.2591}]}.

\bibitem{checkmate2}
D.~Dercks, N.~Desai, J.~S.~Kim, K.~Rolbiecki, \emph{et al}., ``{CheckMATE 2:
  From the model to the limit},''
  \href{https://doi.org/10.1016/j.cpc.2017.08.021}{Comput.\  Phys.\  Commun.\
  {\bfseries 221} (2017) 383--418} {\ttfamily
  [\href{https://arxiv.org/abs/1611.09856}{arXiv:1611.09856}]}.

\bibitem{analysismanager}
J.~S.~Kim, D.~Schmeier, J.~Tattersall, and K.~Rolbiecki, ``{A framework to
  create customised LHC analyses within CheckMATE},''
  \href{https://doi.org/10.1016/j.cpc.2015.06.002}{Comput.\  Phys.\  Commun.\
  {\bfseries 196} (2015) 535--562} {\ttfamily
  [\href{https://arxiv.org/abs/1503.01123}{arXiv:1503.01123}]}.

\bibitem{Baak:2014wma}
M.~Baak, G.~J.~Besjes, D.~C\^ote, A.~Koutsman, \emph{et al}., ``{HistFitter
  software framework for statistical data analysis},''
  \href{https://doi.org/10.1140/epjc/s10052-015-3327-7}{Eur.\  Phys.\  J.\  C
  {\bfseries 75} (2015) 153} {\ttfamily
  [\href{https://arxiv.org/abs/1410.1280}{arXiv:1410.1280}]}.

\end{thebibliography}\endgroup
\bibliographystyle{utphys28mod}

\end{document}